\documentclass[11pt]{article}

\usepackage[margin=2.5cm]{geometry}
\usepackage{hyperref}
\usepackage{float}
\usepackage{verbatim} 
\usepackage{pifont}
\usepackage{multirow}
\usepackage{amsmath,amsfonts,amssymb}
\usepackage{algorithm}
\usepackage{algorithmic}
\usepackage{array}
\usepackage[caption=false,font=normalsize,labelfont=sf,textfont=sf]{subfig}
\usepackage{textcomp}
\usepackage{stfloats}
\usepackage{url}
\usepackage{graphicx}
\usepackage{xcolor}
\usepackage{listings}
\usepackage{subcaption}
\usepackage{caption}
\usepackage{stackengine}
\usepackage{booktabs}
\usepackage[most]{tcolorbox}

\usepackage[authoryear]{natbib}

\makeatletter
\newcommand{\mynote}[2][]{%
  \begingroup
  \tcbset{%
    noteshift/.store in=\mynote@shift,
    noteshift=0.5cm
  }%
  \begin{tcolorbox}[nobeforeafter,
    enhanced,
    sharp corners,
    toprule=1pt,
    bottomrule=1pt,
    leftrule=0pt,
    rightrule=0pt,
    colback=yellow!20,
    #1,
    left skip=\mynote@shift,
    right skip=\mynote@shift,
    overlay={\node[right] (mynotenode) at ([xshift=-\mynote@shift]frame.west) {} ;},
    ]
    #2
  \end{tcolorbox}
  \endgroup
}
\makeatother

\newenvironment{keyword}{
  \vspace{2mm}
  \noindent\textbf{Keywords: } \begin{itshape}
}{
  \end{itshape}\par\vspace{4mm}
}

\newcommand{\sep}{, }

\title{High-level reasoning while low-level actuation in Cyber-Physical Systems: How efficient is it?}

\author{
  Burak Karaduman$^{1,2,*}$,
  Baris Tekin Tezel$^{1,3}$,
  Moharram Challenger$^{1,2}$\\[1ex]
  \small
  $^{1}$Department of Computer Science, University of Antwerp, Middelheimlaan 1, 2020 Antwerp, Flanders, Belgium\\
  \small
  $^{2}$AnSyMo/CoSys Core-lab, Flanders Make Strategic Research Center, 3001 Leuven, Flanders, Belgium\\
  \small
  $^{3}$Department of Computer Science, Dokuz Eylul University, 35390 Buca, Izmir, Turkiye\\[1ex]
  \small
  $^{*}$Corresponding author: \texttt{bburakkaraduman@gmail.com}
}

\date{} 

\begin{document}

\maketitle

\begin{abstract}
The growing complexity of industrial information integration systems requires software technologies that support intelligent behaviour, real-time responsiveness, and efficient development. Despite the proliferation of programming languages and frameworks, there remains a limited amount of empirical evidence to guide engineers in selecting the most suitable tools for developing advanced industrial applications.

This study addresses that gap by measuring and comparing worst-case execution time (WCET) and development time across six languages: C++, Java, Jade, Jason, and fuzzy Jason BDI with loosely and tightly coupled integration. These technologies represent a progression from procedural and object-oriented programming to agent-oriented frameworks that support symbolic and fuzzy reasoning. Instead of relying on broad or ambiguous notions such as paradigms or orientation, we adopt a developer-centred approach based on measurable outcomes. Our structured comparative analysis explores how increasing levels of abstraction and reasoning capabilities influence both the time required to develop applications and their runtime performance. By examining these dimensions, we reveal practical trade-offs among development effort and execution efficiency. Our findings demonstrate how different abstraction levels and reasoning mechanisms influence both system performance and engineering effort. These results provide practical insights for designing intelligent, agent-based systems that operate under real-time constraints and complex decision-making processes.

The study contributes to the ongoing discourse on software selection in industrial informatisation by providing evidence-based guidance that aligns with integration efficiency, software maintainability, and system responsiveness. This work supports future research into the relationship between language features, development dynamics, and runtime behaviour in the context of industrial-oriented cyber-physical and smart manufacturing systems.

\end{abstract}



\section{Introduction}
\label{introduction}

The rapid evolution of programming languages in recent decades reflects the increasing specialization, diversity, and complexity of modern industrial computing systems. These systems often follow layered cloud-fog-edge architectures \citet{gurjanov2021edge}, where each layer operates under specific constraints related to latency, computational resources, and connectivity. Selecting the appropriate abstraction level for each layer is critical for ensuring reliable and efficient integration, thereby increasing system-wide complexity.

Such complexity is particularly pronounced in multi-architecture embedded systems, where developers must operate within stringent conditions such as limited memory, real-time responsiveness, low power consumption, and tight coupling with hardware peripherals. Moreover, the integration of computing, sensing, and communication capabilities has led to the emergence of Cyber-Physical Systems (CPS) and the Internet of Things (IoT), where intelligent decision-making, distributed coordination, and adaptive behavior are key system requirements.

From a combinatorial perspective, IoT and CPS platforms unify physical operations with computation and networked communication \citet{greer2019cyber}. These systems are not only required to respond to environmental stimuli, but also to reason under uncertainty, optimize system performance, and collaborate with other agents or devices across the network. As a result, the demands on software developers have shifted from hardware-level control and optimization to enabling distributed intelligence, autonomy, and context-aware behavior across heterogeneous architectures.

This transformation underscores the need for more expressive and efficient programming languages that support industrial information integration. Developing and maintaining such systems calls for software tools and abstractions that can address both engineering constraints and the cognitive load placed on developers. In this context, understanding the relationship between language design, development effort, and runtime performance becomes crucial for building reliable, intelligent systems in industrial environments.

Such complexity is particularly pronounced in multi-architecture embedded systems, where developers must operate under stringent constraints such as real-time responsiveness, limited memory, low power consumption, and tight coupling with hardware peripherals. Moreover, the landscape of embedded systems has evolved significantly, giving rise to IoT and CPS platforms that extend beyond mere sensor-actuator interaction. These modern systems are characterized by their integration of computing, sensing, communication, and actuation functions, along with a growing demand for distributed coordination, context-awareness, autonomy, and decision-making under uncertainty \citet{greer2019cyber}.

To meet these requirements, programming languages progressively included advanced abstractions that enable programmers to express complicated behaviours more easily. This shift requires programming paradigms that extend beyond hardware interfacing and low-level optimisation.  For example, object-oriented programming (OOP) promotes encapsulation and modularity; functional programming introduces mathematical rigour and immutability; and agent-oriented programming offers autonomy, goals, and reasoning concepts. They aid in handling the complexity of systems by giving conceptual building blocks better suited to the specific problem domain. They tend to come with additional computational overhead or increased execution time, which is a significant detriment to real-time and embedded systems.

Historically, procedural languages like C and C++ have been the default for embedded systems due to their low-level control, minimal runtime overhead, and efficient resource utilization. However, they place a heavy burden on the developer for managing concurrency, memory safety, and modularity. To address these shortcomings, more abstract paradigms emerged over time. Object-Oriented Programming (OOP) introduced encapsulation and reusability through classes and inheritance, while functional programming offered stateless computations, immutability, and higher-order functions for expressing more predictable behaviours.
High-level programming languages, exemplified by Java, incorporate features such as garbage collection, exception handling, and a large set of libraries, thereby enhancing developer productivity at the expense of increased runtime overhead.

Later, the introduction of agent-oriented programming represented a paradigm shift for modelling systems that require autonomy, goals, and reasoning \citet{vachtsevanou2023embedding,
william2022increasing,silva2020embedded}. Agent-based frameworks such as Jade and Jason provide built-in support for high-level constructs like beliefs, desires, intentions, and inter-agent communication. These abstractions are particularly useful in distributed IoT and CPS systems \citet{karaduman2023rational,karaduman2024impact}, where intelligent agents must cooperate, negotiate, and respond adaptively to environmental stimuli and internal objectives.

Moreover, agent-oriented frameworks such as Jade and Jason supply constructs for communication, goal management, and reasoning, thereby enabling systems to model intelligent, autonomous agents \citet{karaduman2023rational}. Fuzzy-BDI (belief-desire-intention) extensions of agent frameworks, such as Jason, have further enhanced their expressiveness by enabling decision-making under uncertainty through the use of fuzzy logic rules. This is particularly beneficial in real-world IoT and CPS contexts, such as robotics, smart environments, or sensor networks, where data is often noisy, incomplete, or ambiguous. By integrating fuzzy logic with agent reasoning, developers can model subtle behaviours such as partial goal satisfaction, graded trust in sensor readings, or prioritisation under vague conditions \citet{karaduman2024impact}. However, these high-level constructs typically come with additional computational complexity and increased overhead, raising questions about their suitability for resource-constrained platforms.

Despite the growing variety of languages and paradigms, there remains a significant lack of empirical evidence comparing them in real-world IoT and CPS development scenarios. Most guidance in the field is based on subjective or loosely defined categories such as “paradigm” or “level of abstraction,” which do little to inform practical decisions. While these categorizations are theoretically helpful, they often fail to account for quantitative engineering concerns like development effort, timing guarantees, or performance bottlenecks. Practitioners, especially those working with embedded, IoT or CPS, are often more concerned with metrics such as WCET, development time, integration complexity, and long-term maintainability than with traditional language characteristics. Additionally, while type systems or syntax are usually the main concerns for theoretical and academic discussions, practitioners are usually interested in quantitative metrics such as development time, performance of execution, scalability, and how easily it integrates with middleware or hardware.

This paper argues for a more systematic and empirical analysis of programming technologies across abstraction levels, ranging from low-level procedural languages to high-level fuzzy-agent-based (hybrid) systems. It presents a comparative evaluation of six representative technologies: C++, Java, Jade, Jason, and two versions of fuzzy-enhanced Jason (loosely and tightly integrated). These platforms were chosen to reflect the spectrum of language abstraction, from bare-metal control to autonomous, uncertainty-aware agents. Each technology has distinct advantages and limitations depending on the use case and deployment environment, particularly in IoT and CPS domains.

The study focuses on two primary evaluation criteria:

\begin{itemize}

\item Worst-Case Execution Time:  a critical metric in embedded IoT and CPS systems where violations of real-time constraints can lead to catastrophic failures. In the relevant studies \citet{wilhelm2008worst,miller2021performance,becker2025expedited,gavigan2024quantifying}, a similar measurement method was used as a basis for various analyses.

\item  Development Time: an indicator of software engineering productivity and effort, which has direct implications for cost, scalability, and maintainability. In the relevant studies, a similar measurement method was used as a basis for various analyses, particularly in evaluating development time comparing various agent languages \citet{kitchenham1996desmet,cossentino2018comparison,adam2017bdi,adam2017comparing,cossentino2018comparison}.

\end{itemize}

Through the implementation of a shared benchmark task in all six technologies, this study offers a quantitative comparison of how abstraction level influences both performance and engineering effort. Furthermore, it seeks to assess whether higher-level paradigms despite their increased expressiveness remain practical and effective within the constraints of IoT and CPS development. The results aim to inform language selection and platform adoption for developers, system architects, and researchers working in domains such as robotics, industrial automation, smart cities, and intelligent sensor networks.

Ultimately, this work attempts to close the persistent gap between the theoretical appeal of high-level programming models and their empirical effectiveness in embedded, hybrid and multi-layer contexts. As IoT and CPS systems grow in scale and complexity, grounded comparisons such as this work are necessary to guide informed decision-making and shape the next generation of development tools and methodologies towards agentic AI \citet{hosseini2025role}.

This paper is organised as follows: Section 2 introduces work related to this study. The required background information for the remainder of this study is provided in Section 3. Methodology is presented in Section 4. Case studies and scenarios are mentioned in Section 5. The empirical and experimental evaluations are provided in Section 6. The paper is discussed in Section 7 and concluded in Section 8.

\section{Related Work}

This section refers to the empirical and experimental studies regarding the aforementioned languages, starting from how we determined them. To decide on languages, we considered the studies in the literature mentioned in this section, major preceding studies \citet{karaduman2021towards,karaduman2022enhancing,karaduman2024impact} and supportive preceding work \citet{ltaief2022agent,yalcin2021agent,schoofs2021software,karnouskos2020industrial}. Therefore, we decided to tackle Jason to program the BDI agents, considering it has been developing for at least two and a half decades \citet{bordini2005jason}. Alternatively, the SPADE-BDI \citet{palanca2020spade} could be preferred,  which is also based on AgentSpeak. For our study, we selected Jason as it runs in the Java environment, which is also suitable for our proposed case studies' compatibility, mentioned in subsection \ref{sec:CS}. 

Study proposes \citet{jordan2015feature} a structured, high-level feature model to compare general-purpose programming languages from the programmer’s perspective, enabling practical comparison across paradigms. They classify and map languages such as C (procedural), Java (object-oriented), Haskell (functional), Erlang (actor-oriented), and Jason (agent-oriented), each representing distinct computational principles. By moving beyond vague notions of paradigms, their model offers a unified basis for systematic language comparison based on feature model representation. Although their detailed feature model-based comparison is comprehensive, they do not provide empirical and WCET analyses.

Study \citet{remmels2017comparison} compares PROFETA (BDI) and  SMACH (state machine) architectures for a collaborative package delivery robot. PROFETA requires less code and uses less real memory (37\% less) but lacks documentation and community support. SMACH, although more complex and memory-intensive, benefits from maturity, improved documentation, and a larger user base. Dynamic performance (CPU, computational time) showed no significant differences. Although PROFETA appears more efficient in implementation, SMACH is found to be more practical due to its usability and support. They emphasise that neither approach is superior; each has trade-offs depending on the specific priorities of the application context. Therefore, this showed us that there is a need for comparing different programming languages from other levels of abstractions, considering development time and WCET measurements based on various case studies.

\citet{naveed2021comparison} compares OOP languages, C++ and Java, based on algorithmic performance. It concludes that all the tests lead to the conclusion that C++ is better than Java in implementing the fundamental algorithms offered in introductory programming courses. In our work, we approach this comparison from the application-oriented view, considering the evolution of the languages towards developing new intelligent methods and corresponding requirements.

Study  \citet{lopes2022and} investigates why complex methods persist in software, despite guidelines encouraging simplicity. It analyzes 1,000 complex methods from 50 open-source projects across Java, JavaScript, Python, C++, and C\#. The results show that complexity often increases over time, especially in C++ and Python. Java and C\# projects show more reduction efforts. A survey with 70+ developers reveals that many do not perceive these methods as problematic. By combining empirical analysis with developer feedback, their study offers a broader view of complexity in practice. It concludes that language affects complexity trends, and developers often keep complex code by choice, not unfamiliarity. In parallel with their findings, our study includes a higher level of abstractions and intelligent approaches to deal with this complexity and explores the suitability of autonomous programming languages for today's systems.


Study \citet{miller2021analysis} performs WCET measurements of Jason on a laptop and RaspberryPi Pi 1 and 4. They found that the average execution times for the reasoning cycles vary significantly depending on the hardware used. The laptop demonstrated the fastest performance with an average execution time of 5,500 microseconds (5.5 milliseconds). Raspberry Pi 4 shows moderate performance with an average of 10,000 microseconds (10 milliseconds), while Raspberry Pi 1 exhibits the slowest execution time at 550,000 microseconds (550 milliseconds). Their results highlight the substantial impact that hardware capabilities have on the processing speed of the system, with newer or more powerful devices executing reasoning cycles much faster. They carried out these experiments on classical Jason experiments and the sole Jason framework. In our study, we specialised each case study for IoT and CPS. We also considered both lower and upper abstractions. Nevertheless, their study inspired us and created a basis for our goal.

 Study \citet{gavigan2024quantifying} examined how design choices in Jason BDI AgentSpeak code impact agent performance and maintainability in a collision avoidance scenario. By applying various software metrics, they found that agents with looser coupling and higher cohesion responded faster and performed better. However, increased complexity led to slower performance due to more intensive rule processing. Focused, smaller plans reduced processing time and improved responsiveness. While rules can slow performance, they enhance maintainability by reducing code duplication. Although well-designed BDI agents outperformed less maintainable ones, they were still slower than equivalent imperative Python programs. Their findings highlight the importance of balancing design to optimise both performance and maintainability. Along with their findings, we conducted measurements by increasing the number of plans and recursive plan calls in the agent part, using both nested if/else statements and flat if/else statements, raising computational complexity. As they inspect development approaches at the Java and Python OOP level.

Study \citet{becker2025expedited} introduces an approach, namely,  EB2A, an expedited BDI agent architecture that reacts at least 2.5 times faster than standard BDI frameworks, maintaining constant response times even under heavy workloads. Their approach preserves intelligent reasoning by adding a lightweight pre-processing layer for critical perceptions, with only about 0.5\% overhead. They integrated it with ROS and validated it in a realistic UAV firefighting scenario. Their solution outperforms pure reflex systems and can benefit from real-time OS and JVM optimisations. Future work may explore handling conflicting actions, prioritisation of critical reactions, and extending to probabilistic BDI frameworks. Such architectural improvements can be added for safety-critical applications. However, our goal is first to inspect how our hybrid approaches work in the worst case during regular operation and their development time efforts, then to consider the response time in an emergency situation.

In a preceding work \citet{karaduman2023rational}, Jason and Jade frameworks were compared both in development time and code complexity for specific development styles, such as implementing Jade agents using sequential and FSM behaviour types and using Jason's reactive and proactive features, resulting in Jason requiring less code complexity and development time. Furthermore,  fuzzy-logic and BDI agents were integrated in the study \citet{karaduman2024impact} in a loosely coupled manner. Based on the primary measurements performed on response time for a single case study, fuzzy logic requires minimal computation time under regular conditions. However, in this study, we examined worst-case execution and development times from both abstraction levels and various use cases.


To synthesise the contributions of prior studies and position our work within the broader research landscape, Table~\ref{tab:literature_comparison} presents a comparative overview of the most relevant literature on programming language evaluation in the context of embedded systems, agent-based development, and cyber-physical systems. While earlier studies have provided valuable insights into language features, execution performance, or software complexity, such as \citet{jordan2015feature} with their structured feature model or \citet{becker2025expedited}  with WCET improvements in BDI agents, few have jointly examined both runtime and engineering effort across multiple paradigms and abstraction levels. As the table illustrates, most existing work focuses on a limited set of languages or evaluation metrics. For instance, \citet{miller2021analysis} analysed WCET for Jason agents, and \citet{gavigan2024quantifying} discussed maintainability trade-offs, but neither considered development effort across multiple agent frameworks. In contrast, our study fills this gap by simultaneously evaluating worst-case execution time, development time, and statistical significance across six distinct technologies, including fuzzy-BDI variants, using multi-case empirical and experimental methods. This holistic perspective distinguishes our work from prior art and aims to guide future platform selection and methodological development in intelligent IoT and CPS design.

\begin{table}[htbp]
\centering
\caption{Comparative Overview of Related Literature}
\label{tab:literature_comparison}
\resizebox{\textwidth}{!}{%
\begin{tabular}{|l|c|c|c|c|c|c|}
\hline
\textbf{Study} & \textbf{Languages Comp.} & \textbf{WCET} & \textbf{Dev. Time} & \textbf{Multi-case} & \textbf{Uncertainty Handling} & \textbf{Stat. Analysis} \\
\hline
\citet{jordan2015feature} & C, Java, Haskell, Jason & No & No & No & No & No \\
\hline
 \citet{remmels2017comparison} & PROFETA vs SMACH & Partial & Partial & No & No & No \\
\hline
\citet{miller2021analysis} & Jason & Yes & No & No & No & No \\
\hline
\citet{becker2025expedited}  & Jason-EB2A & Yes & No & No & Partial & Partial \\
\hline
\citet{gavigan2024quantifying} & Jason, Python & Yes & Partial & No & No & Partial \\
\hline
\citet{naveed2021comparison} & C++, Java & Yes & No & No & No & No \\
\hline
\citet{lopes2022and}& Java, C++, Python, JS, C\# & No & Partial & Yes & No & Partial \\
\hline
\textbf{This Study} & C++, Java, Jade, Jason, Fuzzy-BDI & Yes & Yes & Yes & Yes & Yes \\
\hline
\end{tabular}
}
\end{table}

Ultimately, to the best of our knowledge, it has not been offered both empirical and WCET analyses for various programming languages, starting from OOP to  (fuzzy) AOP ones, in the context of industrial embedded systems, IoT infrastructures, and CPS integration. The following section presents essential background information to provide the necessary context and foundational understanding for the remainder of this paper.

\section{Background}\label{BG}

This section covers the required information about multi-agent systems, Jason BDI agents, simple-reflex Jade agents and fuzzy logic.

\subsection{Multi-Agent Systems}

MAS enhances IoT and CPS  by managing complexity and heterogeneity through autonomy \citet{pico2018agentification} by deploying multiple agents. These agents then collaborate to achieve global objectives, leveraging rational BDI reasoning for decision-making through coordination. MAS emphasises modularity, decentralisation, and reusability, enabling flexible and adaptive operations \citet{merdan2011monitoring}. This way, systems become resilient and adaptive, effectively managing dynamic, unpredictable conditions.  These agents collect and share information and act to solve complex challenges, ensuring IoT and CPS functionality and adaptability across diverse applications while fostering sustainability and efficiency.

\subsection{Jason BDI Agents }

Logic-based reasoning mechanisms, originating in the late 1950s with the development of commonsense reasoning, have evolved to include first-order logic deduction, as well as inductive and abductive reasoning approaches \citet{calegari2021logic}. One of the logic-based models, BDI \citet{bordini2005jason,zhang2022integrated}, relies on its beliefs about the world, which are collected through perceptions and desires that define its objectives. Complex systems, such as logistics and transportation \citet{xu2025multi}, can be modelled, simulated \citet{zhang2022integrated}, and programmed as agents' plans and intentions, which manage their actions to achieve these objectives while considering the current context. Programming languages like Prolog were developed based on these deductive principles. Jason, a Prolog-like agent programming language, integrates the BDI and the Procedural Reasoning System (PRS) architecture, supporting intelligent decision-making       \citet{bordini2007programming}. BDI agents continuously observe and adapt to their environment based on their internal state and context, selecting plans.
\begin{lstlisting}[caption={The form of a Jason plan},label={JasonBDIPlanForm}]
triggering_event: application_context <- plan_body.
\end{lstlisting}

A typical Jason plan follows the structure shown in Listing \ref{JasonBDIPlanForm}. A triggering event (\emph{+!g}), application\_context ($c$), and body ($\mathcal{A}$) constitute of actions $(ac_{i} \in \mathcal{A}$) $(1 \le i \le n)$. A plan body is a series of actions to achieve the desired goal. These actions are carried out through actuators to change the environment, update beliefs and messaging, and tailor the events that occur. The body includes these action sequences and any sub-goals, i.e., goal-decomposition or plan branch-out, directed by the reasoning mechanism. 

\subsection{Simple-reflex Jade Agents}

Jade is a framework designed for agent-based programming that supports the creation of multi-agent systems \citet{bellifemine2000developing}. It also functions as a distributed development platform, offering a flexible infrastructure that can be extended using Java. Jade provides a runtime environment where agents can be instantiated and executed on specific hosts and devices. The agents are constructed at the behavioural level, with each simple-reflex agent comprising various behaviour types tailored to specific goals, tasks, and processes. This framework introduces a behavioural abstraction layer over traditional object-oriented programming, making it well-suited for Java-based environments. Due to these characteristics, we chose to compare two agent development frameworks at different abstraction levels and explore their integration with low-level control mechanisms.

\subsection{Fuzzy Logic}

Zadeh \citet{zadeh1996fuzzy} introduced fuzzy theory to handle imprecise and uncertain information. Fuzzy logic provides an adaptive and comprehensible method of representing rational relationships and enhanced reasoning by dealing with imprecise or vague information. Fuzzy-based systems mimic human thinking. This way, fuzzified systems can better integrate expert knowledge within decision-making processes, expressed in linguistic variables. The essential benefit of this approach is that it can convey sophisticated information and uncertainty, which makes it particularly suitable for modelling complex systems. This enables researchers and practitioners to incorporate human experience and subjective intuition into decision models, resulting in more realistic results, interpretable outputs, and adaptable solutions. Incorporating fuzzy logic into  BDI agents enhances their abilities in decision-making, action execution, reasoning, and plan selection, especially in situations involving run-time uncertainties. This integration contributes to developing more resilient and intelligent complex systems.

A fuzzy set extends the classical set concept, using a membership function to assign each object a degree of membership between zero and one, indicating its belonging to a class. A fuzzy set is a pair $A = (U, m)$ where $U = (-\infty, \infty)$ is described through its membership function $\mu_A : U \rightarrow [0, 1]$, which maps each element of the fuzzy set to a value in the range $[0, 1]$. A fuzzy rule typically follows the structure $R: \text{if } x \text{ is } A, \text{ then } y \text{ is } B$, so A and B are linguistic values represented by fuzzy sets on the discourse universes X and Y.  
The methodology of this study is presented in the following section.

\section{Methodology}\label{methodology}

\begin{figure}[H]
\centering
\includegraphics[width=0.65\columnwidth]{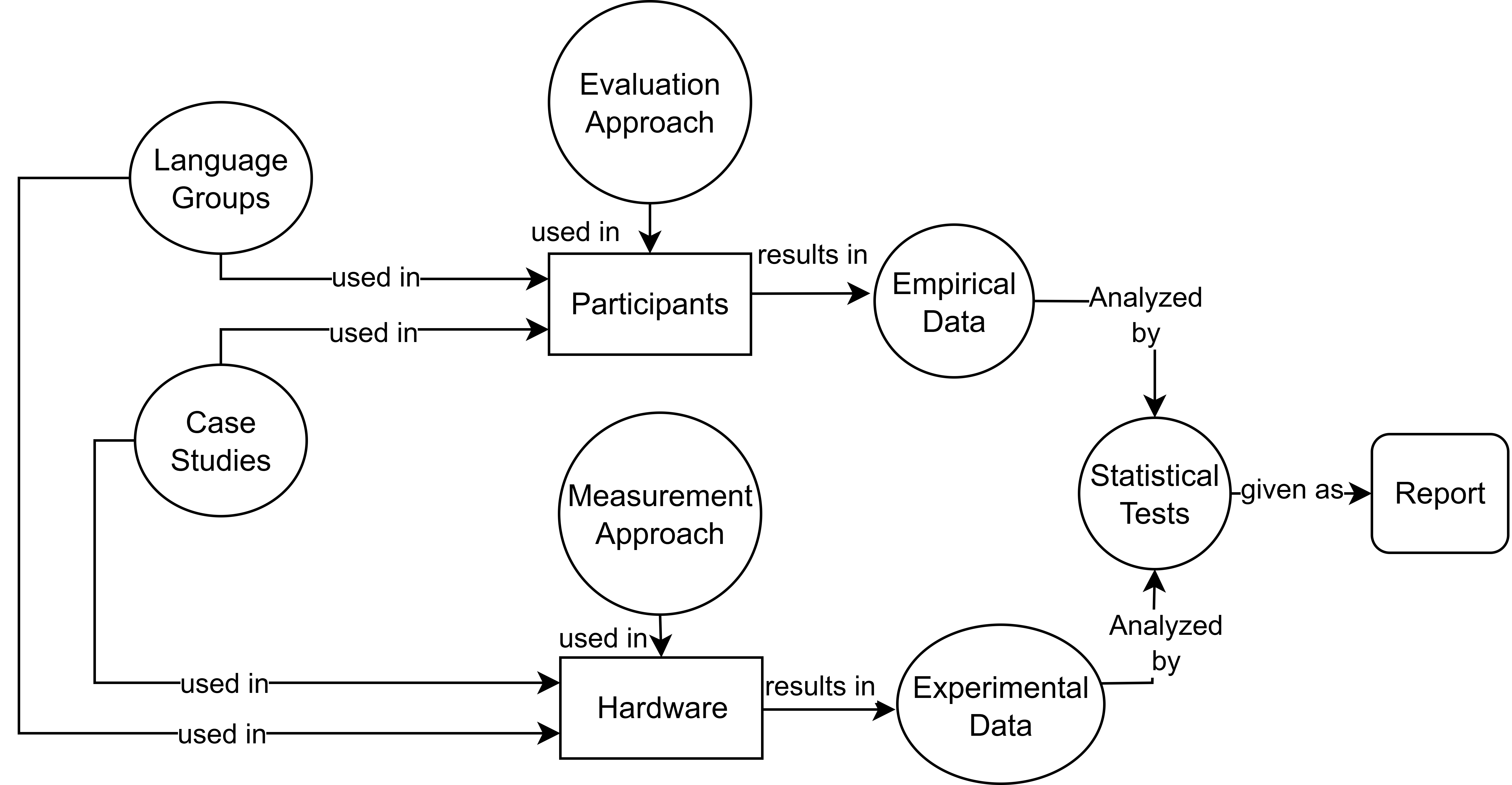}
\caption{High-level overview of the Research Methodology}
\label{fig:ResearchMetholodgy}
\end{figure}

Figure \ref{fig:ResearchMetholodgy} illustrates a conceptual model that describes the overall context in which the research methodology, based on the literature, is derived and utilised. The research methodology is divided into distinct components, including Language Groups, Case Studies, Hardware, Participants, and Evaluation Approach, which collectively guide the empirical and experimental analysis. The  Measurement Approach and Evaluation Approach focus on two key metrics: Worst-Case Execution Time (subsubsection \ref{EmpiricProt}) and Development Time (subsubsection \ref{ExperimProt}).  The Evaluation Language combines empirical data from participant tasks and experimental data from WCET analyses, which are statistically tested to validate findings. The results are synthesised into a comprehensive report (aligning with the high-level overview depicted in Figure \ref{fig:ResearchMetholodgy}.

The study's methodology is systematically structured to evaluate the efficiency of high-level reasoning and low-level actuation in IoT and CPS by comparing six programming technologies: C++, Java, Jade, Jason, and two fuzzy-BDI Jason extensions. The study applies both WCET and Empirical evaluation. For the empirical side, we were inspired by \citep{kitchenham1996desmet} Desmet's methodology. For the experimental side, we were motivated by \citet{wilhelm2008worst} and similar studies \citet{miller2021analysis,miller2021performance,becker2025expedited,gavigan2024quantifying} in the literature.

In the following subsections, the research protocol creation and its application process steps are given.

\subsection{Research Protocol Motivation}

In line with Desmet's framework for technology evaluation in software engineering, this study adopts a mixed-method approach combining both case study and controlled experiment methodologies. Each component of the method was chosen to systematically evaluate the selected technologies in terms of engineering effort and performance. 

\subsubsection{Motivation of Experimental WCET Protocol}

A benchmark scenario was carried out in each language using identical logic, and execution was performed on the same hardware platforms. The primary performance metric, Worst-Case Execution Time, was measured across multiple cycles. Independent language/abstraction type and dependent WCET variables were explicitly defined, and the results were statistically analysed using ANOVA and Tukey HSD tests. This aligns directly with Desmet’s definition of a reproducible and statistically grounded experimental evaluation. In addition, our Desmet-based approach is also aligned with the evaluation methods applied by studies \citet{miller2021analysis,miller2021performance,becker2025expedited,gavigan2024quantifying}, which follow measurement-based analyses from a high-level point of view. In other words, the time interval between input (sense cycle) and output (act cycle) is focused, providing end-to-end coverage. The measurement approach is also mentioned in \citet{wilhelm2008worst}. Specifically, the measurement approach executes the task or task parts on the given hardware for some set of inputs. They then take the measured times and derive the maximal and minimal observed execution times for the whole task, applying end-to-end coverage. Here, we run plan/condition/statements to produce an action based on the given input for each language. 

In \citet{wilhelm2008worst}, it is noted that measurement methods may have limitations in accuracy and are more appropriate for non-hard real-time systems, where static analyses should ideally complement them. Therefore, to provide deeper insights, we analysed the measurements under both realistic and idealised (yet practically unattainable) conditions. To this end, the number of plans of the case studies mentioned in subsection \ref{Scenarios} was increased by a factor of three and ten for the existing cases. It should also be noted that static analyses and simulation approaches are still under research and an open gap \citet{karaduman2024static,karaduman2024towards,tezel2025debugging,xu2023can}.

\subsubsection{Motivation of Empirical Development Time Protocol}

Desmet mentions that tangible impacts are typically measured through reductions in production/development time, rework, or maintenance costs for empirical assessment. Desmet, as a methodology, classifies this type of examination as a quantitative or objective evaluation. Such evaluations involve clearly defining the expected measurable benefits of a new method or tool, and then gathering data to verify whether those benefits are achieved. The appropriateness of a method/tool is usually assessed in terms of the features it provides.  Overall, we follow a multi-case study approach. In this approach, many participants are asked to perform tasks using different programming paradigms. These subjects are carefully established by avoiding bias, and the results can be analysed using statistical techniques. The case studies involve each method being tried out on real projects using standard project development procedures.  To improve the credibility of the outcomes, we expanded both the number of case studies and the participant pool. 

As mentioned in Desmet, Feature Analysis is also a step that involves identifying the requirements participants have for a particular task/activity and mapping those requirements to the features a method/tool should possess to support that task or activity. Therefore, this step is also followed to determine the criteria for the three case studies mentioned in subsection \ref{Scenarios}, considering study \citep{jordan2015feature}. These determined evaluation criteria for the case studies' implementation scenario were attempted to be satisfied with the participants.  The similar methods were also used by studies \citet{junger2016analysis,cossentino2018comparison,adam2017bdi,adam2017comparing,cossentino2018comparison}.

\subsection{Multi-Case Study Protocol for Empirical and Experimental Evaluations}

\begin{figure}[H]
\centering
\includegraphics[width=0.48\columnwidth]{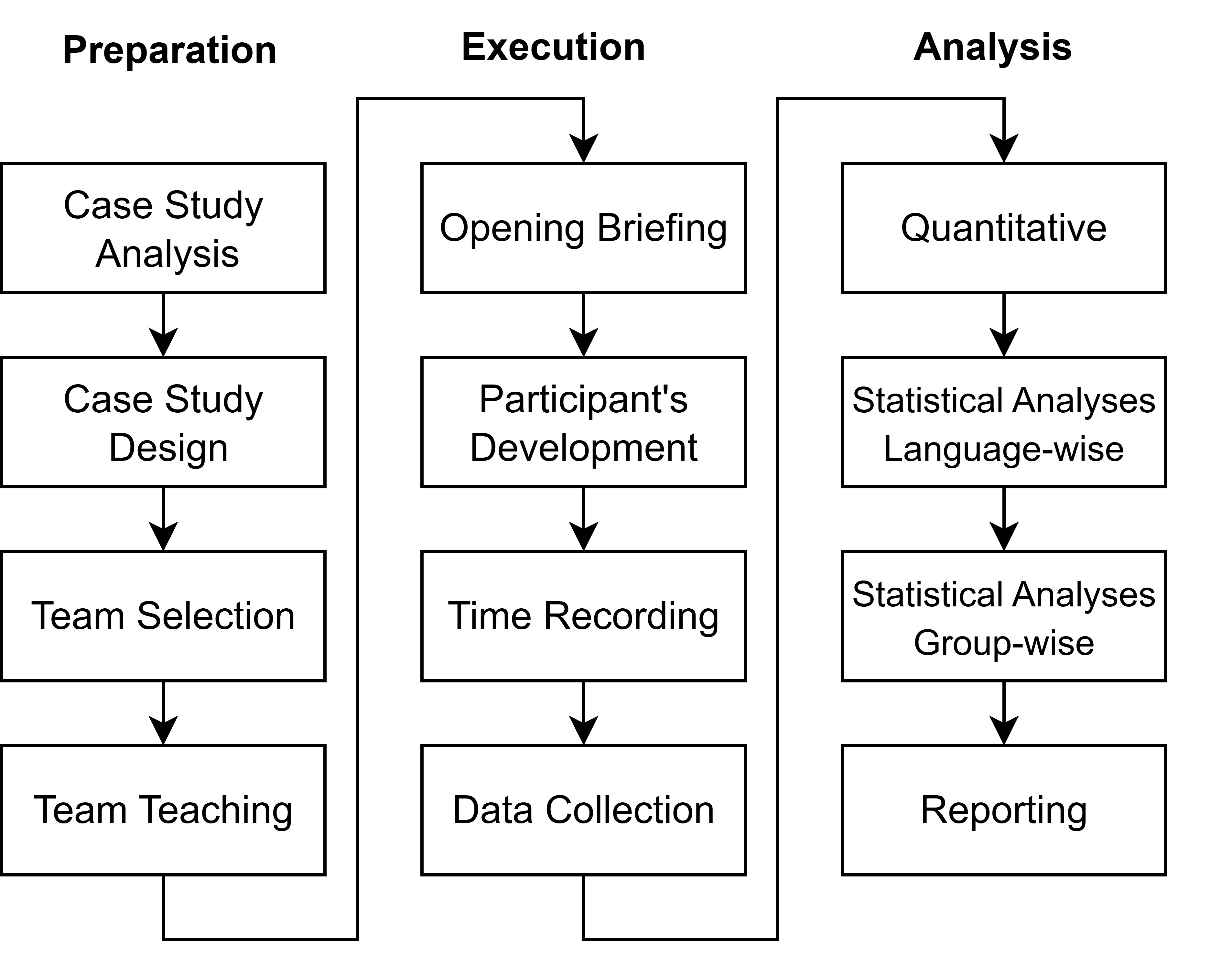}
\caption{The performed process for case study preparation, execution and analyses for empirical evaluation.}
\label{fig:CaseStudyProtocolEmpirik}
\end{figure}

As mentioned, to perform WCET and Empirical analyses, we need to form a protocol for a case study research. In this regard, we consider a three-phase process, which we adapted from \citet{challenger2016systematic}, considering \citet{kitchenham1996desmet}. Figure \ref{fig:CaseStudyProtocolEmpirik} shows the performed steps of preparation, execution and analysis. The case study analysis and design, including their scenarios, are mentioned in subsection \ref{sec:CS} in detail. Team selection, team teaching and briefing are mentioned in subsection \ref{EmpiricProt}. As a summary, the teams are formed as shown by Table \ref{tab:evaluation-groups}. The first group was tasked with implementing two approaches: Jason Fuzzy-BDI Integrated (Tightly Coupled) and Jason Fuzzy-BDI Rule-based, i.e., loosely coupled version. The second group was tasked to implement using Jason Boolean (without any fuzzy enhancement) and Jade languages. Lastly, group 3 performed on Java and C++. The first and second groups consist of 7 MSc students, and the last group has 4 MSc students, 2 BSc students, and 1 PhD student to balance. 

\begin{table}[ht]
\centering
\caption{Empirical Evaluation Group Profiles}
\label{tab:evaluation-groups}
\begin{tabular}{lccc}
\toprule
\textbf{Group} & \textbf{BSc} & \textbf{MSc} & \textbf{PhD} \\
\midrule
Group 1 & 0 & 7 & 0 \\
Group 2 & 0 & 7 & 0 \\
Group 3 & 2 & 4 & 1 \\
\bottomrule
\end{tabular}
\end{table}

While the participants were developing the assigned case studies, we recorded the times for each phase, such as plan/condition/rule writing, deliberation and debugging. We then collected the necessary data and performed quantitative analyses, which are mentioned in subsection \ref{EmpiricalResults}.

\begin{figure}[H]
\centering
\includegraphics[width=0.48\columnwidth]{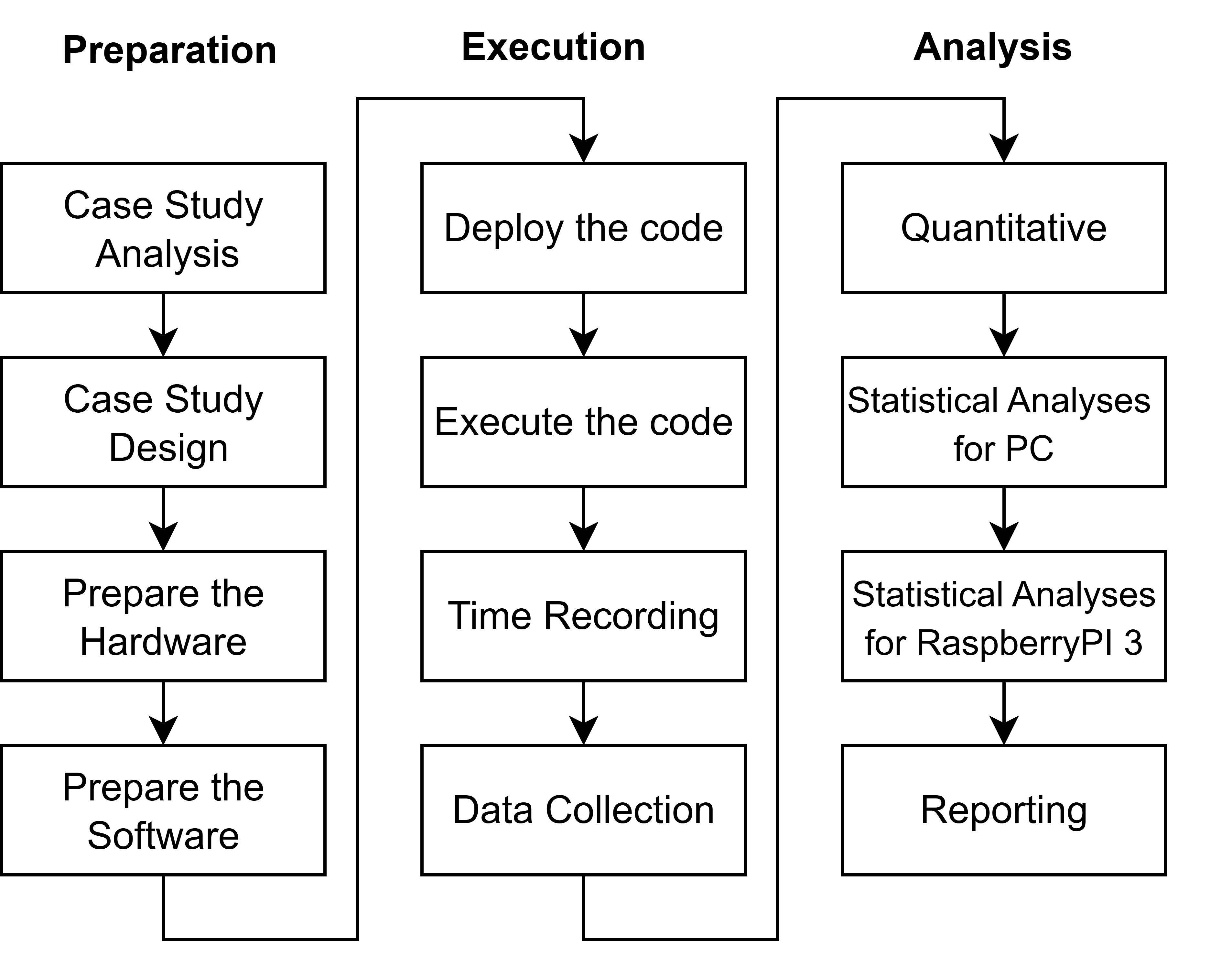}
\caption{The performed process for case study preparation, execution and analyses for experimental evaluation.}
\label{fig:CaseStudyProtocolDeneysel}
\end{figure}

Similarly, Figure \ref{fig:CaseStudyProtocolDeneysel} shows the three-phase experimental evaluation that consists of preparation, execution and analysis. The case study analysis and design phases are the same for the empirical approach. In this part of the study, we prepared the hardware, which consists of both a PC and a Raspberry Pi 3, to run our software. The hardware only ran the codes for our experimental case studies. The software was prepared to run each experimental configuration of every case study for normal and recursive calls, including plan/condition/method multiplication three and ten times. Each software language has a timer library to measure the elapsed time from the beginning of the program until its termination.  This preparation step is mentioned in detail in subsection \ref{ExperimProt}. In the execution phase, the prepared software codes were deployed and executed one by one. The time recordings were automatically taken by the software and written into the log files for both the PC and the Raspberry Pi 3. The data was then collected from the hardware and imported for quantitative analyses. The statistical analyses were performed for both PC and Raspberry Pi 3. The analyses are mentioned in detail in subsection \ref{ExperimResults}. The cross-examination was also performed. The result is then reported.

\subsection{Selected Languages}

C++ is chosen as the foundation language because of its prevalence in embedded systems and performance-sensitive fields. It provides complex hardware manipulation, memory management, and timing that render it appropriate for real-time systems where predictability and low-level manipulation are necessary. Its minimal runtime overhead enables predictable performance, which in  IoT and CPS applications is paramount. However, the language lacks built-in functions that facilitate modularity or high-level reasoning capabilities, which in turn increases the mental effort required of developers and makes it more challenging to create large systems. C++ exemplifies low-level, imperative programming with an emphasis on efficiency, but offers limited abstraction.

Java is included to represent a typical, high-level object-oriented language that provides decent development speed along with modest runtime performance. It features garbage collection, exception handling, multithreading, and an extensive standard library, which reduces the programmer's workload and increases maintainability. Java is used in embedded systems with increased processing capability and at the edge of IoT applications. Its virtual machine-based execution strategy provides portability and platform independence; however, it generally possesses greater runtime overhead than C++. Java achieves a powerful balance between performance and abstraction in both distributed and embedded environments.

Jade is an extension of Java that includes agent-oriented frameworks. It is a formal environment for modeling autonomous agents to share information, coordinate behaviors, and accomplish tasks on a network. Jade adheres to the standards of the Foundation for Intelligent Physical Agents (FIPA) and is widely used in research using multi-agent systems. The platform is especially suited for IoT and CPS applications demanding adaptive communication, distributed control, and adaptive behavior. Bringing agent principles to Java, Jade unites object-oriented and agent-oriented programming and remains compatible with conventional Java tools.

Jason is a more mature variant of agent-oriented programming founded on the BDI reasoning model. Java-based Jason provides special constructs for representing autonomous agents with internal mental states, decision-making strategies, and logic-based plans. Its formal basis in AgentSpeak(L) and support for declarative reasoning make it especially suitable for systems dealing with uncertainty and needing goal-directed behaviour. Jason is seeing growing use in robotics, intelligent automation, IoT and CPS research. It provides a greater level of abstraction than Jade by enabling goal-oriented reasoning rather than merely reactive dialogue.

Jason with fuzzy BDI complements the standard Jason platform by integrating fuzzy logic into the BDI reasoning process. The hybrid model enables agents to handle imprecise input, vague objectives, and partial truths, which is perfect for real-world IoT and CPS scenarios where uncertainty is the rule rather than the exception. For instance, in applications involving uncertain sensor measurements or flexible task ordering, fuzzy-BDI reasoning enables the ability of the system to reach mature, context-dependent judgments instead of relying on strict binary logic. The combination of both loosely coupled and tightly integrated fuzzy extensions illustrates the effect of different degrees of integration on performance, modularity, and complexity for programmers. The various iterations of Jason exhibit the highest degrees of abstraction and reasoning abilities in this research, spanning the spectrum from basic procedural control to sophisticated intelligent autonomy. Briefly, these technologies were chosen carefully to express a wide variety of programming styles and levels of abstraction. With this choice, it is feasible to compare developer effort and execution performance across a spectrum that extends from low-level hardware manipulation to high-level reasoning under uncertain conditions. This comparison is crucial to identify the best language options in designing and building complex IoT and CPS systems.

\subsection{Motivating examples}\label{MotivatingExamples}

In this subsection, motivating examples regarding each paradigm are given. The examples provide control and agent/class/object level details. Any low-level details regarding computation and calculation of sensing/actuation are encapsulated to measure the reflected complexity of high-level requirements fully. Overall, examples follow the sense-decide-act cycle. Firstly, it senses the temperature from the environment using an API, encapsulated by the developer. It then decides which plan/instruction to follow (rule-based system), defined by a low-level design engineer. Lastly, the fan rotation action is executed based on the given parameter. 

Listing \ref{RunningExampleInt} shows a fuzzy-BDI agent code to fuzzy control a fan based on local temperature. A fuzzy-BDI agent program may consist of fuzzy and non-fuzzy parts that can be interoperated.  Line 1 defines an initial goal \emph{sense} (a non-fuzzy plan), which creates a non-fuzzy triggering event. As the only option is the desire written in Line 2, this desire becomes an intention to gather temperature data by \emph{senseEnv} action running the non-fuzzy selectionOption function. The reasoning mechanism then runs \emph{keepCool} goal to create a \textbf{[fuzzy]} triggering event as it is annotated. This way, the options for keepCool event are listed, so option selection is run as fuzzy. Three relevant plans are assessed as options, shown in lines 3-5. The context of each plan is temp(warm), temp(cold) and temp(hot), respectively. The highest membership degree is selected in each agent cycle. From a developer's viewpoint, the programmer must type \textbf{[fuzzy]} annotation for fuzzy plans, actions and triggering events. This is the reflected syntax for the development of integrated fuzzy-BDI agents.

\begin{lstlisting}[caption={Jason Fuzzy-BDI Integrated (Tightly Coupled) Fan Controller},label={RunningExampleInt},otherkeywords={fuzzy,[,],mu,keepCool,:-}]
!sense.
+!sense: true <-  senseEnv; !keepCool[fuzzy]. 
+!keepCool[fuzzy]:temp(warm)<- fanAction(450)[fuzzy].
+!keepCool[fuzzy]:temp(cold) <- fanAction(650)[fuzzy].
+!keepCool[fuzzy]:temp(hot)  <- fanAction(750)[fuzzy].
\end{lstlisting}
 
Lastly, fanAction is run to control the fan using two parameters. The first parameter is given as the maximum action threshold. In other words, a numeric value that the domain expert sets to indicate that the numeric value should be multiplied by the membership degree, which ranges between [0,1]. Therefore, the multiplication result does not exceed this maximum value, proving smooth and fuzzified control. Referring to Listing \ref{RunningExampleInt}, assuming the dominant membership is hot at that moment, the $temp(hot)$ condition is satisfied, and the last plan is selected. The reasoning cycle then tailors the $fanAction(750)$ fuzzy action. Here, 750 is the maximum threshold, representing the fan's highest rotation value when the membership ranges between 0 and 1. 


\begin{lstlisting}[caption={Jason Fuzzy-BDI Loosely Coupled Fan Controller},label={RunningExampleLoose},otherkeywords={fuzzy,[,],mu,:-,keepCool}]
isit(T) :- temp(T,D1) & not(temp(_,D2) & D2>D1).
!init.
+!init: true <- !sample.
+!sample: true <-  sampleSensorData;!keepCool.  
+!keepCool: isit(cold) <- ?temp(cold,D1); fanActionFuzzy(450,D1); !sample.
+!keepCool: isit(warm) <- ?temp(warm,D1); fanActionFuzzy(650,D1); !sample.
+!keepCool: isit(hot)  <-  ?temp(hot,D1); fanActionFuzzy(750,D1);!sample.
\end{lstlisting}

Listing \ref{RunningExampleLoose} shows how loosely coupled fuzzy-BDI agents are formed. In the mentioned integrated form, the programmer has to use $[fuzzy]$ annotation to create fuzzy plans and triggering events. However, loosely coupled fuzzy-BDI agents have relatively more complexity as they are bound to rule definitions, as shown in line 1. The developer initially uses the belief rule on line 1. This rule is used in the plan condition shown in lines 5-7. This rule allows the selection of the plan with the highest membership degree. Unlike the integrated fuzzy-BDI agent form, the highest membership degree is not automatically added to the fuzzy actions in the selected plan. It requires extra instruction,i.e, belief queries such as $ ?temp(cold,D1) $. Here, the highest membership degree is written to $D1$. In the next instruction $fanActionFuzzy(450,D1)$, D1 must be given as a second parameter for multiplication, increasing the complexity. However, this front-end complexity is reduced to using only the $[fuzzy]$ annotation in the integrated fuzzy-BDI version. A relevant reader may find more information on this approach by \citet{karaduman2024impact}.

Listing \ref{RunningExampleBool} lists a conventional (non-fuzzy) BDI agent version. Syntactically, it does not contain any extra instructions as there is no fuzzification (i.e., no uncertainty handling) such as  $[fuzzy]$ annotation, belief rule definition or belief queries. Semantically, it only instructs first-order logic on the background and works with crisp values. Sharp boundaries check the temperature sensing without using membership degrees.  In other words, the temperature values for cold, warm, and hot are always one for one, and zero for the rest of these labels. Therefore, when a plan is selected among 4-6, the fanAction value is directly applied. For example, if plan on line 4 is selected as temp(cold) is one and temp(warm) and temp(hot) are zero, then fanAction is set to 450. This does not provide any enhanced uncertainty handling and only applies traditional BDI-level programming. Our scope is to evaluate the development time and determine the worst-case execution time, comparing the different paradigms.

\begin{lstlisting}[caption={Boolean Fan Controller},label={RunningExampleBool},otherkeywords={fuzzy,[,],mu,:-,keepCool}]
!init. 
 +!init: true <-!sense. 
+!sense: true <-  senseEnv; !keepCool. 
+!keepCool: temp(cold)  <- fanAction(450);!sense.
+!keepCool: temp(warm)  <- fanAction(650);!sense.
+!keepCool: temp(hot)  <- fanAction(750);!sense.
\end{lstlisting}

Listing \ref{RunningExampleJade} shows the Jade implementation of the fan controller example. Jade is an extension of Java that allows using behaviour patterns to deploy simple reflex agents. Initially, the $Agent$ class should be extended to create a Jade agent. Here, a cyclic behaviour (line 3) is utilised as the temperature is sensed in each agent cycle, and the fan speed is set accordingly. An if/else statement can do this fan speed arrangement, as the cyclic behaviour runs continuously.

\begin{lstlisting}[caption={Jade Fan Controller},label={RunningExampleJade},language=Java]
public class FanAgent extends Agent {
    protected void setup() {
        addBehaviour(new CyclicBehaviour(this) {
            public void action() {
                String temp = senseEnv();
                int rpm = getFanSpeed(temp);
                fanAction(rpm);}});}
    private int getFanSpeed(String temp) {
    if (temp.equals("cold"))
    {return 450;}
    else if (temp.equals("warm"))
    {return 650;}
    else if (temp.equals("hot"))
    {return 750;}}}
\end{lstlisting}

Listing \ref{RunningExampleJava} exemplifies the Java version. Here, there is a need for a while loop to mimic the sense-decide-act cycle.  The temperature is sensed by calling the controller instance's sense function (line 5). Then, the label (cold, warm, hot) that satisfies the condition is selected to set the fan speed.

\begin{lstlisting}[caption={Java Fan Controller},label={RunningExampleJava},language=Java]
public class FanController {
    public static void main(String[] args) {
        FanController controller = new FanController();
        while (true) {
            controller.sense(); }}
    public void sense() {
        String temperature = senseEnv();
        keepCool(temperature);}
   public void keepCool(String temp) {
    int speed;
    if (temp.equals("cold"))
    {speed}
    else if (temp.equals("warm"))
    {speed = 650;}
    else if (temp.equals("hot"))
    {speed = 750;}}
    fanAction(speed);}
    private String senseEnv() {...}
\end{lstlisting}

Listing \ref{RunningExampleC++} presents the C++ version of the fan controller. Similar to the Java example, it uses a while loop in the main function to emulate the continuous sense-decide-act cycle. In each iteration, the sense method is invoked on the controller instance (line 16), which first senses the environment (line 3) and then evaluates the temperature label (cold, warm, or hot) inside the keepCool method (lines 8–13). Depending on the condition matched, a corresponding fan speed is set and passed to fanAction for action execution.

\begin{lstlisting}[caption={Boolean Fan Controller (C++)},label={RunningExampleC++},otherkeywords={class,fanAction,while}]
class FanController {
public:
    void sense() {
        std::string temp = senseEnv();
        keepCool(temp); }
private:
    std::string senseEnv() {...}
    void keepCool(const std::string& temp) {
        int rpm = 0;
        if (temp == "cold") rpm = 450;
        else if (temp == "warm") rpm = 650;
        else if (temp == "hot") rpm = 750;
        fanAction(rpm);}
    void fanAction(int rpm) { ...}}};
int main() {
    FanController controller;
    while (true) {
        controller.sense();} return 0;}
\end{lstlisting}


In the following section, requirements for experimental evaluation and implementation details are mentioned.

\section{Case Studies and Scenarios}

In this section, we introduce the case studies and goals that the participants were asked to satisfy. The participants were presented with three case studies on CPS, IoT, and House Robotics, each illustrating different agent and class use variants. 

\subsection{Case Studies}\label{sec:CS}

In this subsection, three case studies related to various domains are introduced. 

\subsubsection{Network Scaler: A Self-Adaptive IoT Case Study }

\begin{figure}[H]
\centering
\includegraphics[width=0.45\columnwidth]{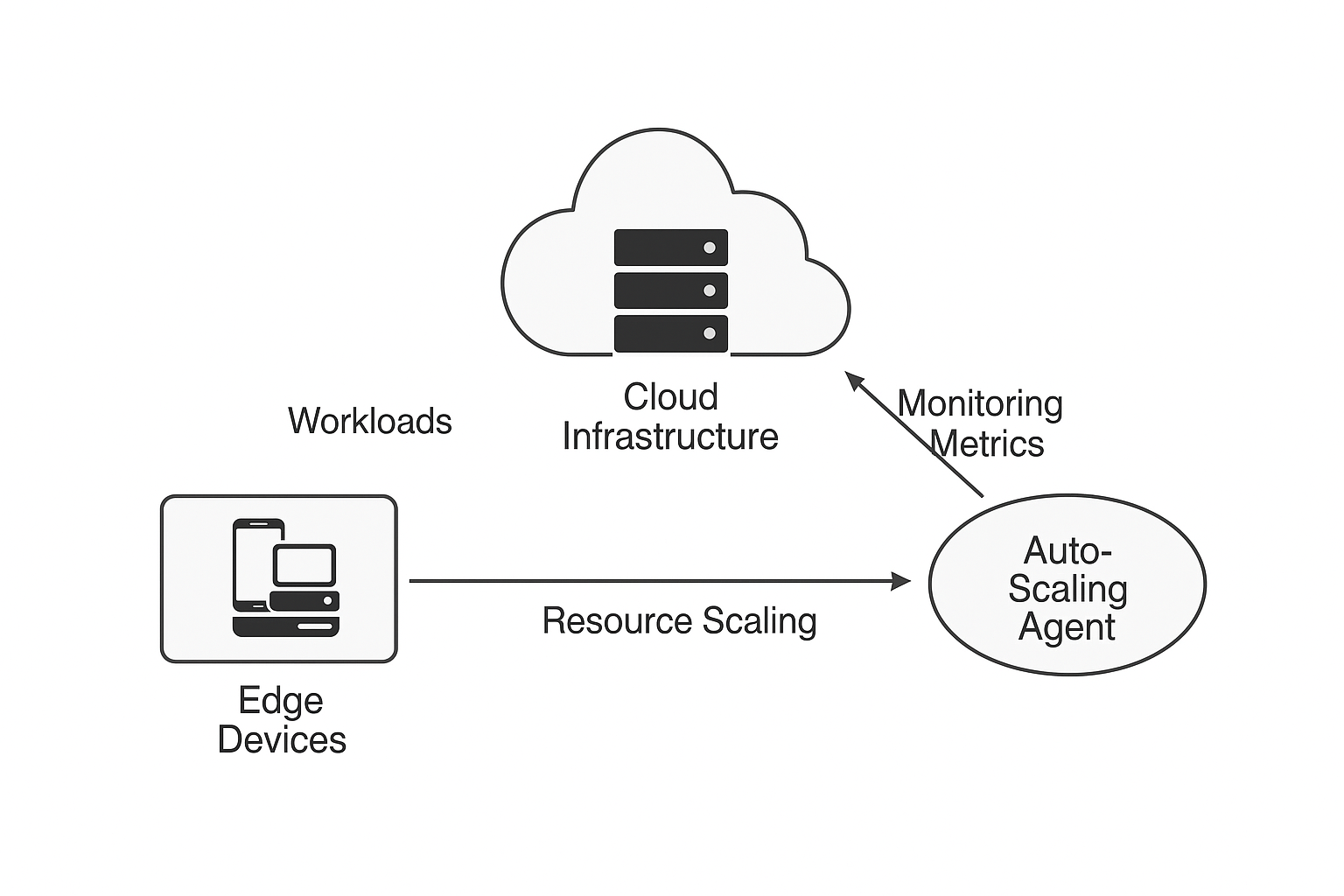}
\caption{Network Scaler System.}
\label{fig:NSConcept}
\end{figure}

Adjusting from the case study in \citet{weyns2020introduction}, it is suggested a cloud infrastructure that minimises costs amid dynamic and irregular workloads based on auto-scaling. Such cloud services align with edge-level devices, such as mentioned \citet{iftikhar2017deltaiot,provoost2019dingnet}. Whilst linear correlations between performance metrics and resource assignments are common, they often fall short in complex scenarios. Defining effective rules can be challenging due to limited infrastructure knowledge and operating conditions. Therefore, a rule-based approach offers a way to correlate scaling factors, adapting to varying workloads and response times. This method, as shown by Figure \ref{fig:NSConcept} aims to balance the limited resources in fluctuating environments. In particular, these rules are formed to minimise cloud computing while consuming the workload using as few computers as possible. In short, single class/agent runs for increasing/decreasing workload and decides to add or deduct computing resources to consume the upcoming workloads.

Nine rules, shown in Listing \ref{lst:NetworkScalerFuzzyRules},  are reflected as if/else statements or (fuzzy) plans and are asked to be implemented by the participants.  The plans/statements use two beliefs/belief rules (workLoad and responseTime), and two actions (scale factor and consumeWorkLoad). The system calculates response time by $total workload/numberOfMachines$ on each iteration. These kinds of implementations are abstracted away from the participants and labelled by linguistic terms such as low workload and good response time. The participants were only focused on the high-level details, such as writing the control/reasoning/rule logic. shows the control rules to be implemented by the participants. A relevant demo can be accessed on \footnote{\url{https://www.youtube.com/watch?v=-SO42lizn88}} to get insights.

\begin{lstlisting}[caption={Rule specification for fuzzy plan definition.},label={lst:NetworkScalerFuzzyRules},otherkeywords={IF,IS,AND,THEN,;},basicstyle=\tiny]
IF workLoad IS low AND responseTime IS good THEN scaleFactor(-15);
IF workLoad IS low AND responseTime IS ok THEN scaleFactor(-10);
IF workLoad IS low AND responseTime IS bad THEN scaleFactor(10);
IF workLoad IS medium AND responseTime IS good THEN scaleFactor(-10);
IF workLoad IS medium AND responseTime IS ok THEN scaleFactor(0);
IF workLoad IS medium AND responseTime IS bad THEN scaleFactor(10);
IF workLoad IS high AND responseTime IS good THEN scaleFactor(0);
IF workLoad IS high AND responseTime IS ok THEN scaleFactor(10);
IF workLoad IS high AND responseTime IS bad THEN scaleFactor(15);
\end{lstlisting}

\subsubsection{Cleaning Robots: A House Robotics Case Study}

This case study was altered for our purpose from the version which had been used in study \citet{becker2025expedited}. In this modified version, two cleaning robots are controlled by 27 established rules, shown in Listing \ref{lst:CleaningRobotsRules}.   The robots have constrained resources, such as vacuum bag capacity and battery power. Therefore, robots arrange the vacuum power according to these rules. A system engineer establishes these rules, and battery thresholds and vacuum bag limits are already set. The participant's role is to convert these rules to the target language and satisfy the requirements in the provided document.

\begin{figure}[H]
\centering
\includegraphics[width=0.40\columnwidth]{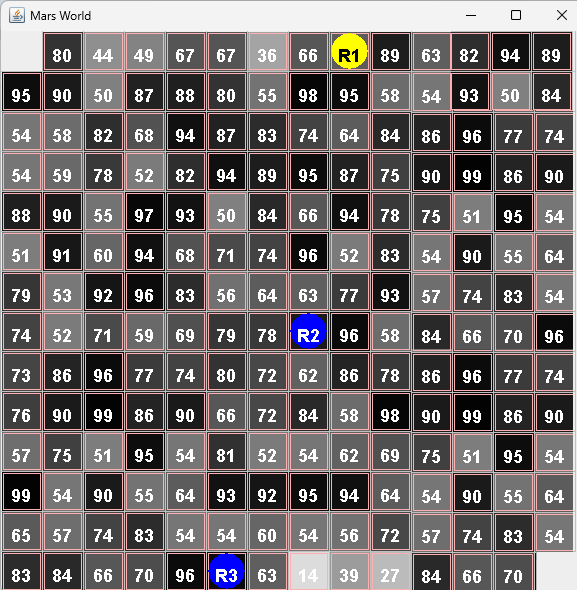}
\caption{Cleaning Robots Grid-map.}
\label{fig:CleaningRobotsGrid}
\end{figure}

Dirt intensity (ranges 0.0-100.0) on each tile, shown by Figure \ref{fig:CleaningRobotsGrid}. Each robot starts cleaning at the opposite corners and is responsible for half of the map. However, when a robot finishes its area and reaches the middle of the map, it tells the other robot to continue cleaning the other's area. While cleaning their zones, they should monitor their vacuum capacity and battery power. This approach enables robots to adjust vacuum power dynamically based on the current tile's capacity, battery level, and dirt intensity, optimising cleaning efficiency and resource use. The robots' objective is not to clean the entire grid but to intelligently manage their resources and make the space appear fairly homogeneous and clean. A demo related to this case study can be found at \footnote{\url{https://youtu.be/rroAxiCWTbo}}.

\begin{lstlisting}[caption={Rule specification for fuzzy plan definition.},label={lst:CleaningRobotsRules},otherkeywords={IF,IS,AND,THEN,;},basicstyle=\tiny]
   IF batteryPower IS min AND vacuumBag IS min AND dirtIntensity IS mid THEN burnGarb(70);
   IF batteryPower IS min AND vacuumBag IS min AND dirtIntensity IS max THEN burnGarb(90);
   IF batteryPower IS min AND vacuumBag IS mid AND dirtIntensity IS min THEN burnGarb(50);
   IF batteryPower IS min AND vacuumBag IS mid AND dirtIntensity IS mid THEN burnGarb(70);
   IF batteryPower IS min AND vacuumBag IS mid AND dirtIntensity IS max THEN burnGarb(90);
   IF batteryPower IS min AND vacuumBag IS max AND dirtIntensity IS min THEN burnGarb(50);
   IF batteryPower IS min AND vacuumBag IS max AND dirtIntensity IS mid THEN burnGarb(70);
   IF batteryPower IS min AND vacuumBag IS max AND dirtIntensity IS max THEN burnGarb(90);
   IF batteryPower IS mid AND vacuumBag IS min AND dirtIntensity IS min THEN burnGarb(50);
   IF batteryPower IS mid AND vacuumBag IS min AND dirtIntensity IS mid THEN burnGarb(70);
   IF batteryPower IS mid AND vacuumBag IS min AND dirtIntensity IS max THEN burnGarb(90);
   IF batteryPower IS mid AND vacuumBag IS mid AND dirtIntensity IS min THEN burnGarb(50);
   IF batteryPower IS mid AND vacuumBag IS mid AND dirtIntensity IS mid THEN burnGarb(70);
   IF batteryPower IS mid AND vacuumBag IS mid AND dirtIntensity IS max THEN burnGarb(90);
   IF batteryPower IS mid AND vacuumBag IS max AND dirtIntensity IS min THEN burnGarb(50);
   IF batteryPower IS mid AND vacuumBag IS max AND dirtIntensity IS mid THEN burnGarb(70);
   IF batteryPower IS mid AND vacuumBag IS max AND dirtIntensity IS max THEN burnGarb(90);
   IF batteryPower IS max AND vacuumBag IS min AND dirtIntensity IS min THEN burnGarb(50);
   IF batteryPower IS max AND vacuumBag IS min AND dirtIntensity IS mid THEN burnGarb(70);
   IF batteryPower IS max AND vacuumBag IS min AND dirtIntensity IS max THEN burnGarb(90);
   IF batteryPower IS max AND vacuumBag IS mid AND dirtIntensity IS min THEN burnGarb(50);
   IF batteryPower IS max AND vacuumBag IS mid AND dirtIntensity IS mid THEN burnGarb(100);
   IF batteryPower IS max AND vacuumBag IS mid AND dirtIntensity IS max THEN burnGarb(100);
   IF batteryPower IS max AND vacuumBag IS max AND dirtIntensity IS min THEN burnGarb(70);
   IF batteryPower IS max AND vacuumBag IS max AND dirtIntensity IS mid THEN burnGarb(100);
   IF batteryPower IS max AND vacuumBag IS max AND dirtIntensity IS max THEN burnGarb(100);
\end{lstlisting}

\subsubsection{Smart Production Line: A CPS Case Study}

\begin{figure}[H]
\centering
\includegraphics[width=0.55\columnwidth]{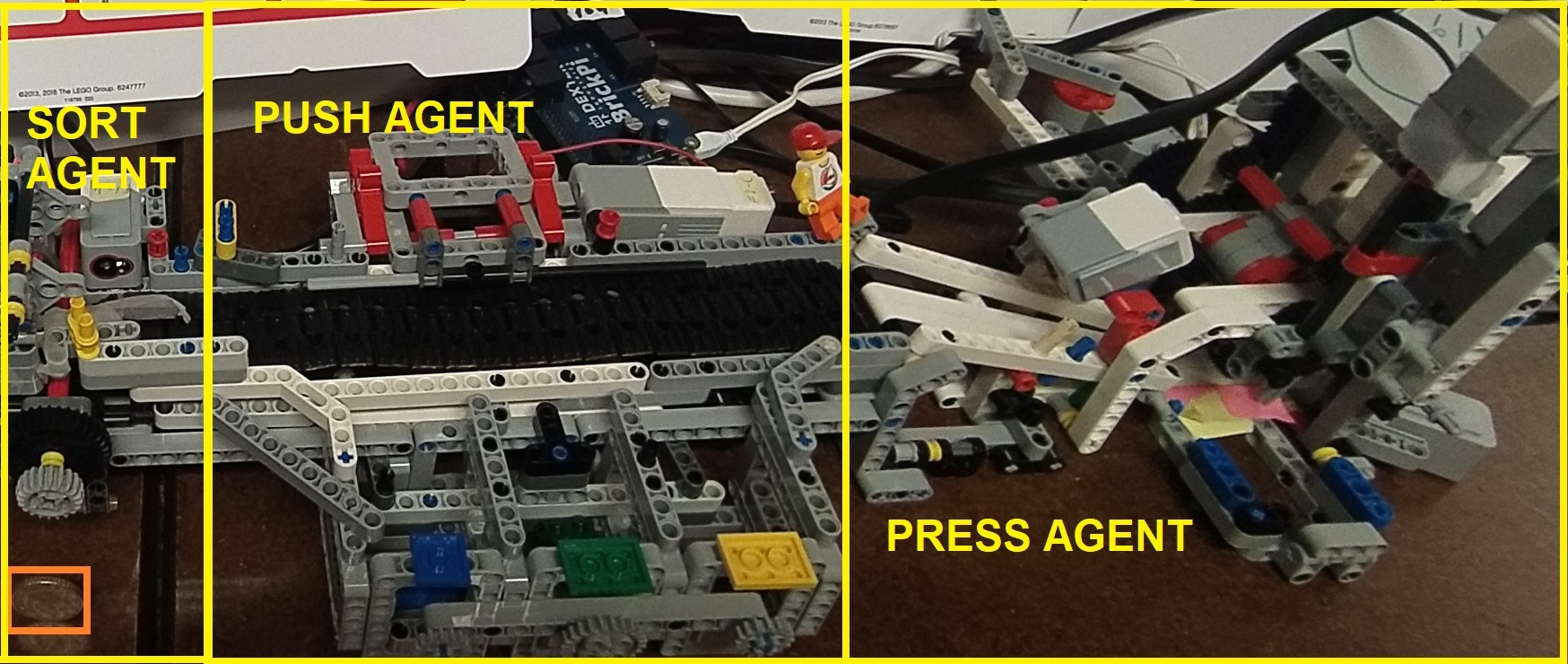}
\caption{Smart production line system.}
\label{fig:SPLHalf}
\end{figure}

This case study is similar to the one used in \citet{karaduman2024impact}. We borrowed it for this work as it is complex. Figure \ref{fig:SPLHalf} illustrates the smart production line case study. The system has 3 agents/classes: Sort, Push, and Build. The rule-based process control focuses on the Sort, which uses sensory data to determine the colour of products, such as tomatoes (red bricks) and peppers (green bricks). In other words, rules consider the R, G, and B colour combination. Moreover, R, G, and B variations for each colour combination of a product should be combined within certain thresholds. These thresholds were labelled such as low red, middle green, high blue, etc, and encapsulated by a system engineer. The participant only implements the rules written on the requirements sheet.

The Sort applies 36 rules, shown in Listing \ref{lst:SmartProductionLineRules}, to sort and control products defined by the stakeholder. Unwanted products are sent to the Push section, and wanted ones are kept by sending them to the Build section. The Push moves the products out of the production line. In our scenario, the stakeholder only wants the tomato sauce, i.e., accepting only red bricks. The Build presses the tomatoes to create sauce. Sample videos of this case study can be found on \footnote{\url{https://www.youtube.com/watch?v=SmSpZaDiBSI}} and \footnote{\url{https://www.youtube.com/watch?v=tWh5w6eEABQ}}.

\begin{lstlisting}[caption={Rule specification for Smart Production Line System.},label={lst:SmartProductionLineRules},otherkeywords={IF,IS,AND,THEN,;},basicstyle=\tiny]
IF red IS high AND green IS low AND blue IS medium THEN Red
IF red IS high AND green IS medium AND blue IS medium THEN Red
IF red IS high AND green IS high AND blue IS low THEN Red
IF red IS medium AND green IS medium AND blue IS low THEN Red
IF red IS medium AND green IS high AND blue IS low THEN Red
IF red IS medium AND green IS veryhigh AND blue IS low THEN Red
IF red IS high AND green IS veryhigh AND blue IS low THEN Red
IF red IS medium AND green IS medium AND blue IS medium THEN SpoiledRed
IF red IS medium AND green IS high AND blue IS medium THEN SpoiledRed
IF red IS high AND green IS high AND blue IS medium THEN SpoiledRed
IF red IS medium AND green IS medium AND blue IS high THEN SpoiledRed
IF red IS high AND green IS medium AND blue IS low THEN SpoiledRed
IF red IS medium AND green IS veryhigh AND blue IS medium THEN SpoiledRed
IF red IS medium AND green IS veryhigh AND blue IS high THEN SpoiledRed
IF red IS high AND green IS veryhigh AND blue IS high THEN SpoiledRed
IF red IS medium AND green IS ultramedium AND blue IS medium THEN Light Green
IF red IS medium AND green IS ultralow AND blue IS medium THEN Light Green
IF red IS medium AND green IS ultrahigh AND blue IS medium THEN Light Green
IF red IS medium AND green IS ultramedium AND blue IS high THEN Light Green
IF red IS medium AND green IS ultrahigh AND blue IS high THEN Light Green
IF red IS high AND green IS ultramedium AND blue IS medium THEN Light Green
IF red IS high AND green IS ultramedium AND blue IS high THEN Light Green
IF red IS high AND green IS ultrahigh AND blue IS medium THEN Light Green
IF red IS high AND green IS ultrahigh AND blue IS high THEN Light Green
IF red IS low AND green IS ultralow AND blue IS medium THEN Middle Green
IF red IS low AND green IS ultralow AND blue IS medium THEN Middle Green
IF red IS low AND green IS ultralow AND blue IS medium THEN Middle Green
IF red IS low AND green IS ultralow AND blue IS medium THEN Middle Green
IF red IS low AND green IS veryhigh AND blue IS medium THEN Middle Green
IF red IS low AND green IS veryhigh AND blue IS medium THEN Middle Green
IF red IS low AND green IS veryhigh AND blue IS high THEN Middle Green
IF red IS low AND green IS ultralow AND blue IS high THEN Dark Green
IF red IS low AND green IS high AND blue IS low THEN Dark Green
IF red IS low AND green IS high AND blue IS medium THEN Dark Green
IF red IS low AND green IS veryhigh AND blue IS low THEN Dark Green
IF red IS low AND green IS low AND blue IS low THEN Dark Green
\end{lstlisting}

\subsection{Scenarios}\label{Scenarios}

In this subsection, the scenarios of the case studies, their requirements and code excerpts are given.

\subsubsection{Scenario 1: Network Scaler}\label{subsec55}

In this experiment, the participants were asked to implement the given rules. The participants were not directed to any development method or implementation approach. The first fuzzy-BDI group is required to implement the given rules mentioned in Listing \ref{lst:NetworkScalerFuzzyRules} as Jason plans. Integrated version (lines 1-12) requires $[fuzzy]$  annotation as shown in Listing \ref{lst:NetworkScalerFuzzySolution}, lines 4-12. The loosely coupled version (lines 15-28) requires additional belief rule writing as shown in lines 15-16. The loosely coupled version must use these belief rules to determine the highest membership degree for plan selection. One belief rule is required for each belief used by a fuzzy plan. 

For both integrated and loosely coupled versions, the agent requires an initial goal that can be named as \emph{start or initial}, lines 1 and 17, respectively. The participants should also continuously create a sense-deliberate-act convention. At the beginning of the \emph{start} plan, the agent receives a new workload and then branches out to the \emph{!scale} fuzzy plan. The participants should also branch out between plans by defining event triggers such as \emph{!scale}. 

The participants do not begin the experiments from scratch, but by a given template where the agents/classes are related to low-level details. The template contains a small piece of agent/class codes, such as in lines 2-3 and 19-20, to trigger workload and response time calculations in the background.

This is needed as the network scaler agent should sense when to receive a new workload (lines 2-3, 18-19), decide on workload size, calculate the response time, and select a suitable plan accordingly (4-12, 20-28). Lastly, the plan chosen arranges the scaling up or down of the system resources, i.e., the number of computers, instructing \emph{scaleFactor} and \emph{consumeWorkload} actions. Eventually, the network scaler agent should perform these steps continuously until the internal mechanism stops the application. 

\begin{lstlisting}[caption={A Sample solution for Group 1's the Fuzzy-BDI Network Scaler case.},label={lst:NetworkScalerFuzzySolution},otherkeywords={start,+!,!,[fuzzy],<-,scale,scaleFactor,scaleFactorF,isitworkLoad,isitresponseTime},basicstyle=\tiny]
!start.  // Integrated Version
+!start : (currentWorkLoad(CWL) & CWL==0) |(arrivedTurn(AVV) & AVV==0)<-getWorkLoadBool;checkWorkLoad;!scale[fuzzy];!start.
+!start : currentWorkLoad(CWL) & CWL\==0  <-checkWorkLoad;!scale[fuzzy];!start.
+!scale[fuzzy]: workLoad(low) & responseTime(good) <-    scaleFactorF(-15)[fuzzy]; consumeWorkLoad.
+!scale[fuzzy]: workLoad(low) & responseTime(ok) <-      scaleFactorF(-10)[fuzzy]; consumeWorkLoad.
+!scale[fuzzy]: workLoad(low) & responseTime(bad) <-   scaleFactorF(10)[fuzzy]; consumeWorkLoad.
+!scale[fuzzy]: workLoad(medium) & responseTime(good) <- scaleFactorF(-10)[fuzzy]; consumeWorkLoad.
+!scale[fuzzy]: workLoad(medium) & responseTime(ok) <- scaleFactorF(0)[fuzzy]; consumeWorkLoad.
+!scale[fuzzy]: workLoad(medium) & responseTime(bad) <- scaleFactorF(10)[fuzzy]; consumeWorkLoad.
+!scale[fuzzy]: workLoad(high) & responseTime(good) <-scaleFactorF(0)[fuzzy]; consumeWorkLoad.
+!scale[fuzzy]: workLoad(high) & responseTime(ok) <-scaleFactorF(10)[fuzzy]; consumeWorkLoad.
+!scale[fuzzy]: workLoad(high) & responseTime(bad) <- scaleFactorF(15)[fuzzy]; consumeWorkLoad.
...
// Loosely Coupled Version 
isitworkLoad(Is) :- workLoad(Is,S1) & not(workLoad(_,S2) & S2>S1).
isitresponseTime(Is) :- responseTime(Is,S1) & not(responseTime(_,S2) & S2>S1).
!start.
+!start: (currentWorkLoad(CWL) & CWL==0) |(arrivedTurn(AVV) & AVV==0)<-getWorkLoadBool;checkWorkLoadRule; !scale;!start.
+!start: currentWorkLoad(CWL) & CWL\==0  <- checkWorkLoadRule;!scale;!start.
+!scale: (isitworkLoad(low) & isitresponseTime(good)) <-scaleFactor(-15); consumeWorkLoad.
+!scale: (isitworkLoad(low) & isitresponseTime(ok)) <- scaleFactor(-10); consumeWorkLoad.
+!scale: (isitworkLoad(low) & isitresponseTime(bad))  <-scaleFactor(10); consumeWorkLoad.
+!scale: (isitworkLoad(medium) & isitresponseTime(good))  <-scaleFactor(-10); consumeWorkLoad.
+!scale: (isitworkLoad(medium) & isitresponseTime(ok))  <-scaleFactor(0); consumeWorkLoad.
+!scale: (isitworkLoad(medium) & isitresponseTime(bad))  <-scaleFactor(10); consumeWorkLoad.
+!scale: (isitworkLoad(high) & isitresponseTime(good))  <-scaleFactor(0); consumeWorkLoad.
+!scale: (isitworkLoad(high) & isitresponseTime(ok))  <-scaleFactor(10); consumeWorkLoad.
+!scale: (isitworkLoad(high) & isitresponseTime(bad))  <-scaleFactor(15); consumeWorkLoad.
\end{lstlisting}

For the conventional agent group, Listing \ref{lst:NetworkScalerGroup2} shows the traditional Jason and Jade sample solutions written as non-fuzzy. The requirements are more or less the same here. For this Jason version (lines 1-12), the participants do not need to use any specific annotation or belief rule as there is no fuzzy-logic enhancement \citet{karaduman2024impact}. Lines 14-45 show the Jade version. The rules can be implemented as if/else statements (lines 14-35). A cyclic behaviour of a Jade network scaler agent can run continuously (lines 37-45).

\begin{lstlisting}[caption={A Sample solution for Group 2's the Fuzzy-BDI Network Scaler case.},label={lst:NetworkScalerGroup2},otherkeywords={start,+!,!,[fuzzy],<-,scale,scaleFactor,scaleFactorF,isitworkLoad,isitresponseTime},basicstyle=\tiny]
!start.
+!start: (currentWorkLoad(CWL) & CWL==0) |(arrivedTurn(AVV) & AVV==0) <-getWorkLoadBool;checkWorkLoadBool;  !scale; !start.
+!start: currentWorkLoad(CWL) & CWL\==0 <-checkWorkLoadBool; !scale;!start.
+!scale: (workLoad(low) & responseTime(good))<-scaleFactor(-15); consumeWorkLoad.
+!scale: (workLoad(low) & responseTime(ok))  <-scaleFactor(-10); consumeWorkLoad.
+!scale: (workLoad(low) & responseTime(bad)) <-scaleFactor(10); consumeWorkLoad.
+!scale: (workLoad(medium) & responseTime(good))  <-scaleFactor(-10); consumeWorkLoad.
+!scale: (workLoad(medium) & responseTime(ok))   <-scaleFactor(0); consumeWorkLoad.
+!scale: (workLoad(medium) & responseTime(bad))  <-scaleFactor(10); consumeWorkLoad.
+!scale: (workLoad(high) & responseTime(good))  <-scaleFactor(0); consumeWorkLoad.
+!scale: (workLoad(high) & responseTime(ok))   <- scaleFactor(10); consumeWorkLoad.
+!scale: (workLoad(high) & responseTime(bad))  <- scaleFactor(15); consumeWorkLoad.
...
void scale(ScalerAgentEnvironment sae )
{int factor = 0;
if (workLoad.equals("low") && responseTime.equals("good")) {
factor = -15;
} else if (workLoad.equals("low") && responseTime.equals("ok")) {
factor = -10;
} else if (workLoad.equals("low") && responseTime.equals("bad")) {
factor = 10;
} else if (workLoad.equals("medium") && responseTime.equals("good")) {
factor = -10;
} else if (workLoad.equals("medium") && responseTime.equals("ok")) {
factor = 0;
} else if (workLoad.equals("medium") && responseTime.equals("bad")) {
factor = 10;
} else if (workLoad.equals("high") && responseTime.equals("good")) {
factor = 0;
} else if (workLoad.equals("high") && responseTime.equals("ok")) {
factor = 10;
} else if (workLoad.equals("high") && responseTime.equals("bad")) {
factor = 15;}    
sae.scaleFactor(factor);
sae.consumeWorkLoad();}
....
 addBehaviour(new CyclicBehaviour(this) {
 @Override
 public void action() { try { if ((currentWorkLoad==0)||arrivedTurn==0) {
sae.getWorkLoadBool();
sae.checkWorkLoadBool();
this.arrangeResourceScale();
} else if (currentWorkLoad!=0) {
sae.checkWorkLoadBool();
this.arrangeResourceScale();}
\end{lstlisting}

Lastly, as shown in Listing \ref{lst:NetworkScalerGroup3}, for OOP Group 3, it requires a function which contains these rules (lines 1-22). A \emph{while} loop is required to run the agent/class continuously. Lastly, an object instantiation is required for the network scaler agent class.  For brevity, we only show the Java version, but the C++ version is also nearly the same for this case study. The details can be reached on \footnote{\url{https://github.com/micss-lab/ExperimentFiles/tree/main/sampleSolutions}}

\begin{lstlisting}[caption={A Sample solution for Group 3's the OOP Network Scaler case.},label={lst:NetworkScalerGroup3},otherkeywords={start,+!,!,[fuzzy],<-,scale,scaleFactor,scaleFactorF,isitworkLoad,isitresponseTime},basicstyle=\tiny]
void scale(ScalerAgentEnvironment sae )
{int factor = 0;
if (workLoad.equals("low") && responseTime.equals("good")) {
   factor = -15;
} else if (workLoad.equals("low") && responseTime.equals("ok")) {
   factor = -10;
} else if (workLoad.equals("low") && responseTime.equals("bad")) {
   factor = 10;
} else if (workLoad.equals("medium") && responseTime.equals("good")) {
   factor = -10;
} else if (workLoad.equals("medium") && responseTime.equals("ok")) {
   factor = 0;
} else if (workLoad.equals("medium") && responseTime.equals("bad")) {
   factor = 10;
} else if (workLoad.equals("high") && responseTime.equals("good")) {
   factor = 0;
} else if (workLoad.equals("high") && responseTime.equals("ok")) {
   factor = 10;
} else if (workLoad.equals("high") && responseTime.equals("bad")) {
   factor = 15;}    
sae.scaleFactor(factor);
sae.consumeWorkLoad();}
while (true) {
 if ((currentWorkLoad == 0) || arrivedTurn == 0) {
sae.getWorkLoadBool(); sae.checkWorkLoadBool();
arrangeResourceScale(sae);
} else if (currentWorkLoad != 0) {
sae.checkWorkLoadBool();arrangeResourceScale(sae);}}
\end{lstlisting}

To summarise:
\begin{enumerate}
    \item The nine rules shall be written as plans/conditions.
    \item Continuous loop/plan branching shall be satisfied.
    \item Required syntax and annotation shall be used, if necessary.
\end{enumerate}

 In the following subsection, the cleaning robots experiment, its requirements and scenario details are mentioned.

\subsubsection{Scenario 2: Cleaning Robots}\label{subsec56}

In this case study, the participants were asked to implement the given 27 rules for two robots. The rules consider three parameters/beliefs/belief rules: battery power, vacuum bag capacity and dirt intensity and one action (burnGarb) for each robot. For integrated fuzzy-BDI Jason, the participants must use $[fuzzy]$ annotation, while the loosely-coupled fuzzy-BDI Jason version requires three belief rules, as shown in Listing \ref{lst:CleaningRobotsGroup1} (lines 1-3), as there are three control parameters.  Additional goals that must be implemented by the participants are some collaborative messaging between the agents and their coordinates on the grid map. Robot 1 starts at the topmost left corner (point 0,0) while Robot 3 is initiated at the bottommost right corner (point 13,13). Robot 2 is located at point 7,7 for coordination. 

On the given requirement (question) document, they were asked to satisfy three goals: i) \emph{check(slots)} plan must continuously run to move a robot until it reaches the middle of the map (see R2 on Figure \ref{fig:CleaningRobotsGrid}). At the same time, the robot's vacuum bag must not be full, or the battery cannot be depleted. For Robot 1, these are points 6,7, while it is 8,7 for Robot 3. When the robot moves, it should trigger a (fuzzy) plan/function such as \emph{!arrangeVacuumPower}. This plan/function contains 27 rules to clean the current tile based on the control parameters (line 5), applying the \emph{burnGarb} action/function. However, if the robot arrives at its destination, such as 6,7 or 8,7, then it must inform other cleaning robots by sending a message such as \emph{.send(r3,tell,continue(r3,true))}.  At the same time, it should also inform Robot 2. The cleaning mission continues unless the robot 2 receives  $moveCount(\_,2)$ from both robots. This messaging varies for a few situations. ii) Any robot may arrive at its destination with enough resources to move on (line 9). Therefore, the cleaning mission can continue, and the cleaning robot also sends $moveCount(\_,1)$  to the coordinator robot 2. Any cleaning robot's resources may be exhausted by arriving at the destination. If it cannot continue, it sends a continue message to the other cleaning robot and $moveCount(\_,2)$ to the coordinator robot (line 11). iii) It can arrive at the destination, but at the same time, any robot resource gets depleted. When both cleaning robots cannot continue because of depleted resources, they send $moveCount(\_,2)$ to the coordinator Robot 2 (line 7). The mission/application finished. Therefore, the participants shall implement these conditions. The given template only contains the plan/condition at line 5 (also for the other languages). Lastly, the participants should modify line 5 and add \emph{$|$continue(\_,true)}. This is needed for the robots' continuation after they meet in the middle of the map (6,7 and 8,7). If they have enough resources, the plan at line 5 is triggered to move the robots, which have enough resources to continue, towards the end corners of the map until their resources get exhausted. These goals shall be written for two cleaning robot agents/instances as a homogeneous case study.

To summarise:
\begin{enumerate}
    \item The  27 rules shall be implemented as plans/conditions.
    \item Three message sending states shall be devised.
    \item Continuous loop/branching should be satisfied.
    \item Robot's coordinates shall be determined correctly.
    \item The robots shall stop once they reach the middle of the map.
    \item If a robot cannot reach the middle, it signals to continue to the other robot.
    \item If a robot receives a continue message from the other, it continues to the end of the map, not the middle.
    \item A robot can only continue as long as its resources allow it to.
    \item Given plan's/condition's expression shall be modified by $|$continue(\_,true).
    \item Required syntax and annotation shall be used, if necessary.
    \item These requirements shall be met for two robots.
\end{enumerate}

Overall, Group 1 shall consider $[fuzzy]$ annotation to implement integrated Jason's fuzzy plans (line 14) while loosely coupled Jason fuzzy plans require belief rule definitions (lines 1-3) used in fuzzy plans' condition shown in line 15.

\begin{lstlisting}[caption={A Sample solution excerpt for Group 1's fuzzy-BDI Cleaning Robots case.},label={lst:CleaningRobotsGroup1},otherkeywords={start,+!,!,[fuzzy],&,check(slots),AgentR3,.,continueSignal,|},basicstyle=\tiny]
isitbatteryPower(Is) :- batteryPower(Is,S1) & not(batteryPower(_,S2) & S2>S1).
isitvacuumBag(Is) :- vacuumBag(Is,S1) & not(vacuumBag(_,S2) & S2>S1).
isitdirtIntensity(Is) :- dirtIntensity(Is,S1) & not(dirtIntensity(_,S2) & S2>S1).
....
+!check(slots):  ((not pos(r1,6,7)) | continue(r1,true)) & vacuumBagFull(r1,M) & M==empty & batteryCharge(r1, BR1) & BR1==full
<-next(slot);checkStatus;!arrangeVacuumPower[fuzzy];!check(slots).
+!check(slots) : ( pos(r1,6,7) & ((vacuumBagFull(r1,ML) & ML==full) | ( batteryCharge(r1, BR1) & BR1==depleted))  ) 
 <-.send(r3,tell,continue(r3,true)); .send(r2,tell,moveCount(r1,2)).  //;  R1 can NOT move further, yet arrives right on the 6,7 and has finished its area.
+!check(slots) : (   & pos(r1,6,7) & ((vacuumBagFull(r1,MM) & MM==empty) | ( batteryCharge(r1, BR1) & BR1==full))  ) 
 <-  .send(r3,tell,continue(r3,true)); .send(r2,tell,moveCount(r1,1)).  //;  R1 can move further, yet arrives it has finished its area.
+!check(slots) : ((vacuumBagFull(r1,MK) & MK==full) | ( batteryCharge(r1, BR1) & BR1==depleted)) 
 <-.send(r3,tell,continue(r3,true)); .send(r2,tell,moveCount(r1,2)).  //;    R1 cannot move any further.
...
+!arrangeVacuumPower[fuzzy]: (batteryPower(min) & vacuumBag(min) & dirtIntensity(min))  <-  burnGarb(50,1).
+!arrangeVacuumPower: (isitbatteryPower(min) & isitvacuumBag(min) & isitdirtIntensity(min))  <- burnGarb(50,1).
 
\end{lstlisting}

For brevity and simplicity, the code excerpts were shortened. In Group 2, the traditional Jason version requires no belief rule creation and $[fuzzy]$ annotation as shown in Listing \ref{lst:CleaningRobotsGroup2}, line 3. Lines 1 and 2 show that the conventional Jason version branches out to \emph{arrangeVacuumPower} plan without annotation. The plan sample on line 3 also does not use any belief rule. Conventional Jason relatively requires less syntactic complexity. For the Jade, Java and C++ versions, these messaging mechanisms were asked to implement as boolean flags such as \emph{continue = true}. This is requested to inspect the complexity of the implementations/experiments at the control logic level. Participants shall also modify the if statement with \emph{AgentR3.continueSignal} and \emph{AgentR1.continueSignal} to allow two robots to continue cleaning after they meet in the middle of the map. They can only continue to move on if they have enough resources. The messaging states and system control rules were expected to be implemented as if/else statements mentioned in Listing \ref{lst:CleaningRobotsRules}. For the Jade version, a CyclicBehaviour, shown in Listing \ref{lst:CleaningRobotsGroup2}, lines 5-15, is used with if/else statements to satisfy messaging, arrange vacuum power for cleaning, and meet the listed requirements. The remaining languages' sample implementations can be found in this repo \footnote{\url{https://github.com/micss-lab/ExperimentFiles/tree/main/sampleSolutions}}

\begin{lstlisting}[caption={A Sample solution excerpt for Group 2's traditional agents Cleaning Robots case.},label={lst:CleaningRobotsGroup2},otherkeywords={start,+!,!,[fuzzy],&,check(slots)},basicstyle=\tiny]
+!check(slots): (not pos(r1,6,7) | continue(r1,true)) & vacuumBagFull(r1,M) & M==empty & batteryCharge(r1, BR1) & BR1==full
<- next(slot);checkStatusBool; !arrangeVacuumPower;!check(slots).
+!arrangeVacuumPower: (batteryPower(min) & vacuumBag(min) & dirtIntensity(min))  <-  burnGarb(50,1).
...
addBehaviour(new CyclicBehaviour(this) {
 @Override
public void action() {
try { if (((x != 6 || y != 7) || AgentR3.continueSignal) && !batteryPower.equals("depleted") && !vacuumBag.equals("full")) {
agent1.checkStatus();
agent1.arrangeVacuumPower("Ag1");
agent1.nextSlot();
} else if (batteryPower.equals("depleted") || vacuumBag.equals("full")) {                   
continueSignal = true;}
if (x == 6 && y == 7 && !batteryPower.equals("depleted") && !vacuumBag.equals("full")) {     
continueSignal = true;}   }}});
\end{lstlisting}

In the following subsection, the smart production line experiment, its requirements and scenario details are mentioned.




\subsubsection{Scenario 3: Smart Production Line}\label{subsec57}

In a smart production line scenario, borrowed from \citet{karaduman2023rational}, we have partitioned three agents/classes: Sort, Push, and Build. This way, we included a heterogeneous case study where all the system portions have different roles and functionalities. This case study also demonstrates the interoperability between fuzzy-BDI and conventional BDI. The participants were tasked with finding the intentionally injected bugs and completing the incomplete parts. The Sort applies the control logic and has 36 rules, as listed in \ref{lst:SmartProductionLineRules}, to decide on the incoming products via a conveyor belt. The Sort agent has three beliefs/belief rules. One action for printing the result of a product's colour. The Build agent uses one belief. 

The Sort Agent sends the red product to the Build section and the non-reds to the Push section. The Push section receives a message, typically a text from the Sort section, specifically \emph{push}. This appears as a simple internal message passing in Jason versions using the \emph{.send} internal action and function calls in the Jade, Java and C++. The reason is that we wanted to abstract away the message communication and network engineering details between agents/classes from the participants as much as possible, focusing on measuring control, behaviour, and interaction level functionalities. 

The participants shall implement the correct messaging keyword to establish interaction between these sections. The Push section shall receive the \emph{push} keywords, and the Build section shall receive the \emph{build}. When the Push section receives a message from the Sort section, it pushes the product out by calling the \emph{pushProduct function/action}. This way, we measure how the participants can read and interpret the provided code and fix the bugs. The Build section performs state-machine-like behaviour. At first, it receives a red product, it increases a counter, namely, \emph{buildStatus}. If counter is one, \emph{pressOnce} function/action is called. It presses the red product once. When it receives another red product for the second time, it increases the \emph{buildStatus} counter to two. When the counter is two, it then calls \emph{pressTwice} function. It presses the red products to merge them. It then calls the eject function to eject the products outside the system and sets the counter to zero, returning to the initial state. Lastly, the Sort section applies 36 control rules, shown in Listing \ref{lst:SmartProductionLineRules}, to sort or push the unwanted product outside the production line.

To summarise:
\begin{enumerate}
    \item The 36 rules shall be implemented as plans/conditions for the Sort section.
    \item Message sending functionalities shall be devised among the three agents.
    \item Continuous loop/branching should be satisfied for the Sort section.
    \item State-chart-like behaviour should be satisfied for the Build section.
    \item Reactive behaviour shall be provided for the Push section, as it acts when it receives a message.
    \item Intentional bugs should be resolved for the Push section.
    \item The red products shall be led to the Push section to be pushed.
    \item The non-red products shall be directed to the Build section to be harnessed. 
    \item Required syntax and annotation shall be used, if necessary.
    \item These requirements shall be met for three sections.
\end{enumerate}

In Listing \ref{lst:SmartProductionLineGroup1}, a sample solution excerpt for the integrated Jason fuzzy-BDI version is shown in Lines 1-12. Sort agent samples the colour in line 1 and branches out to the \emph{decidecolorF} plan. In the background, the product's colour is decided based on fuzzy functions and membership degrees. However, this is abstracted from the participants. The participants shall implement a way to instruct the 36 rules mentioned in \ref{lst:SmartProductionLineRules} according to the sampled colour and decided colour labels. Two excerpts of these rules are shown in lines 1-2. As the product's colour is determined as red, then the Sort agent sends it to the Build section; otherwise, it dispatches the product to the Push section. The Push agent runs the plan in line 7 when it receives a message to push the product outside. If the Build agent receives a red product and message, then it instructs lines 9-12 to realise the aforementioned state-chart-like behaviour. It should be noted that the Sort agent is a fuzzy-BDI agent while the Build and Push agents are BDI agents. This is a feature provided by our framework, namely, FAIR. This allows both to interoperate and implement both agent types and provide code re-use for a heterogeneous case study. This way, the participants of Group 1 did not repeat the developments for the Push and Build agents. For Group 2, they performed the conventional Jason programming of these agents and their Jade version, which can be reached on \footnote{\url{https://github.com/micss-lab/ExperimentFiles/tree/main/sampleSolutions}}. The Jade version required a cyclic behaviour for each agent to send and receive messages, as it is one of the implementation patterns \citet{bellifemine2001developing}. Lines 14-25 show a sample Build Agent, and lines 27-37 exhibit a sample Push Agent implementation in Jade. Lastly, Group 3 preferred to implement these three sections as classes using OOP approaches, if/else statements and function calls for messaging. The rest of the sample implementations for all the target languages can be found on \footnote{\url{https://github.com/micss-lab/ExperimentFiles/tree/main/sampleSolutions}}

\begin{lstlisting}[caption={A sample solution excerpt for Group 3's fuzzy-BDI agents Smart Production Line case.},label={lst:SmartProductionLineGroup1},otherkeywords={start,+,!,[fuzzy],&},basicstyle=\tiny]
+!samplecolor: true  <- sampleColor; !decidecolorF[fuzzy]; !samplecolor.
+!decidecolorF[fuzzy]: (red(high) & green(low) & blue(medium))  <-   saveResult("Red");!toBuild.
+!decidecolorF[fuzzy]: red(medium) & green(ultramedium) & blue(medium)    <-  saveResult("LightGreen"); !toPush.
+!toPush: true <-.send(pushAgent,achieve,push). 
+!toBuild: true <- .send(buildAgent,achieve,build).
...
+!push : true <- pushMotor.
...
+!build : true <- -+buildFree(false); buildStatus(M); K = M+1; -+buildStatus(K); !state.
+!state : buildStatus(1)<- PressOnce. 
+!state : buildStatus(2)<- PressTwice; !eject
+!eject : true <- -+buildStatus(0); ejectProduct.
...
protected void setup() {
addBehaviour(new CyclicBehaviour() {
 public void action() {
 ACLMessage msg = receive(); // Receive a message
 if (msg != null)
 {if ("build".equalsIgnoreCase(String.valueOf(msg.getContent()))) { buildStatus++;
 checkState(); }} else {block();}}); }});}
 private void checkState() {
 if (buildStatus == 1) {
 } else if (buildStatus == 2) {
 eject();}}
 private void eject() {buildStatus = 0;}
 ...
addBehaviour(new CyclicBehaviour() {
public void action() {
ACLMessage msg = receive(); // Receive a message
if (msg != null) {
if ("push".equalsIgnoreCase(String.valueOf(msg.getContent()))) {
              addBehaviour(new OneShotBehaviour() {
              public void action() {
                pushProduct();}});}
               } else { block(); }}});
public static void pushProduct() {
System.out.println("The Product is Pushed away."); }
 \end{lstlisting}

In the following section, the experimental evaluations and timings for each group are given.

\section{Evaluation}\label{Eval}

This section covers empirical and experimental evaluations.  In the following subsections, the results of empirical and experimental measurements are given after their application protocols are presented. The gathered data as input and the analysis output can be found on the GitHub repo, respectively \footnote {\url{https://github.com/micss-lab/ExperimentFiles/tree/main/input}} and \footnote{\url{https://github.com/micss-lab/ExperimentFiles/tree/main/output}}.

\subsection{Empirical Evaluation Protocol}\label{EmpiricProt}

The empirical experiments conducted in this study were designed to be unbiased and methodologically balanced. A total of 21 participants were involved, evenly distributed into three groups (7 participants each). Each group carried out the experiments using two different programming environments, selected from the following: Jason BDI with Fuzzy integration, Jason-BDI Fuzzy Rule, Jason Boolean, JADE, Java, and C++. Participants were carefully chosen to be similar in terms of technical background, experience level, and programming knowledge. This ensured that comparisons between groups were fair and valid. The participants were informed through a 10-minute briefing session to introduce the cases and requirements, and were taught by 1-hour educational videos to refresh their memories, just in case. In addition, before starting their first case study, they were asked to implement the motivating examples mentioned in subsection \ref{MotivatingExamples} for the corresponding languages assigned to them. They then began to develop the case studies.

Their development process was monitored to measure engineering effort, primarily through recording the total development time and observing the planning, coding, and debugging phases. The combination of realistic scenarios, participant-based execution, and time-based measurement makes this setting consistent with Desmet’s case study structure.

Each group implemented three different scenarios using two distinct programming languages. The language assignments were balanced across groups to minimise potential learning effects or biases arising from familiarity with a particular language. No participant was assigned to more than one group, thereby avoiding any strategy transfer or information contamination between groups.

The first group consisted of individuals with at least one year of experience in Jason BDI and relevant industrial exposure. Participants in this group were from Belgium, Turkiye, and Ireland, and were selected such that they had no prior acquaintance with one another. This helped reduce the risk of intra-group bias due to personal familiarity. The second group was composed of academic and industrial developers from Belgium and Turkiye, again with participants selected to ensure no prior relationship. The third group consisted of participants from the research groups of different universities, representing a variety of academic disciplines. This diversity was intended to mitigate potential institutional or educational background bias.

The experiments were conducted in two sessions. In the first session, participants implemented the three scenarios in one programming language, having two 15-minute breaks. In the second session, they implemented the same scenarios using the alternate language. A 10-day break was intentionally introduced between the two sessions. This interval was designed to reduce cognitive bias, especially learning transfer, memory of specific implementations, and fatigue effects, allowing participants to approach the second session with a more neutral mindset.

Several measures were taken to preserve the integrity of the experiments. Depending on availability, participants joined the sessions either remotely or in person. Remote sessions were monitored via screen sharing to ensure adherence to the rules. No communication or information sharing was permitted between groups, and participants who knew each other were not placed in the same group. All groups were provided with the same pre-recorded instructional video as a warm-up and orientation material. Finally, all collected data were anonymised, and the evaluation process was carried out independently of the participants’ group affiliations.

\subsection{Experimental Evaluation Protocol}\label{ExperimProt}

The experiment of this study was carried out by performing a measurement-based approach on each case study. To carry out WCET analyses, we targeted measurements between sensing and acting during the experimental evaluation, including the decision step. We aimed to determine the computational cost and conduct a worst-case analysis, starting from the sensing phase and proceeding to deliberation and action. The measurements regarding the WCET analyses were performed on the aforementioned realistic case study using a PC and a Raspberry Pi 3 B+ with a BrickPi extension board \citet{karaduman2023rational,miller2021analysis,miller2021performance}.

The PC used for the experiments had 16 GB of RAM and an Intel(R) Core(TM) i7-10850H processor with a base frequency of 2.70 GHz and a maximum frequency of 2.71 GHz. We compared six languages. As an approach, we placed a timer just before the sensing from the environment was performed and stopped the timer when an action was executed. We initially measured the condition/state and plan selection based on the aforementioned rule size in cases 9, 27, and 36, respectively. We then increased the plan size by three and ten times. For the sake of brevity and to avoid repetition, we do not list the x3 and x10 multiplication of each case study. This experimental preparation involved copying and pasting the same rules, which pertained multiple times, to be reasoned and executed during runtime.  Lastly, we measured the nested plan and if/else calls for each language abstraction. This nested structure began with a fully nested format, then transitioned to a flat-first and nested-second style, and ultimately reverted to a nested-first and flat-second format for each given rule of each case study.

Listing \ref{lst:ProductionLineWCET} shows how that nested call was performed for the smart production line case study. As can be seen, lines 2-17 indicate that the first portion, which consists of 18 plans, is written as a nested plan, while the remaining rules, in lines 18-23, are flat for the Sort agent, section, and class for the other languages. Lines 26-29 demonstrate how it was set up for Jade, Java, and C++. The reason for including only the Sort agent is that it has the highest fuzzy-BDI form for comparison with the other language abstraction levels. The same approach was applied to two agents/instances of the Cleaning Robots case study and the Network Scaler study.

\begin{lstlisting}[caption={Network Scaler WCET experiments structure.},label={lst:ProductionLineWCET},otherkeywords={start,+!,!,[fuzzy],&,check(slots),AgentR3,.,continueSignal,|},basicstyle=\tiny]
+!samplecolor: true  <- sampleColor; !decidecolorF1[fuzzy]; !samplecolor.
+!decidecolorF1[fuzzy]:(red(high) & green(low) & blue(medium))<-saveResult("Red");
!decidecolorF2[fuzzy].  
+!decidecolorF2[fuzzy]:(red(high) & green(medium) & blue(medium))  <-saveResult("Red");
!decidecolorF3[fuzzy]. 
+!decidecolorF3[fuzzy]:(red(high) & green(high) & blue(low))
<-saveResult("Red");
!decidecolorF4[fuzzy]. 
+!decidecolorF4[fuzzy]:(red(medium) & green(high) & blue(low))     <-saveResult("Red");
!decidecolorF5[fuzzy]. 
+!decidecolorF5[fuzzy]:(red(medium) & green(medium) & blue(low))   <-saveResult("Red");
!decidecolorF6[fuzzy]. 
+!decidecolorF6[fuzzy]:(red(medium) & green(veryhigh) & blue(low)) <-saveResult("Red");
!decidecolorF7[fuzzy].
+!decidecolorF7[fuzzy]:(red(high) & green(veryhigh) & blue(low))   <-saveResult("Red");
!decidecolorF8[fuzzy].
...
+!decidecolorF18[fuzzy]:(red(medium) & green(ultrahigh) & blue(medium))<-saveResult("Light Green");!decidecolorF[fuzzy].
+!decidecolorF[fuzzy]: (red(medium) & green(ultramedium) & blue(high))<-saveResult("Light Green").
+!decidecolorF[fuzzy]: (red(medium) & green(ultrahigh) & blue(high))<-saveResult("Light Green").
+!decidecolorF[fuzzy]: (red(high) & green(ultramedium) & blue(medium))<-saveResult("Light Green").
+!decidecolorF[fuzzy]: (red(high) & green(ultramedium) & blue(high))<-saveResult("Light Green").
+!decidecolorF[fuzzy]: (red(high) & green(ultrahigh) & blue(medium))<-saveResult("Light Green").
+!decidecolorF[fuzzy]: (red(high) & green(ultrahigh) & blue(high) )<-saveResult("Light Green").
...
if ((isRed.equals("high") && isGreen.equals("low") && isBlue.equals("medium")) || true) {
            if ((isRed.equals("high") && isGreen.equals("medium") && isBlue.equals("medium")) || true) {             
if ((isRed.equals("medium") && isGreen.equals("ultramedium") && isBlue.equals("medium")) || true) {....   }... nested structure continues...}}}}}}}}}}}}}}}}}
//  flat if/else statements for 18-36
if ((isRed.equals("medium") && isGreen.equals("ultralow") && isBlue.equals("medium"))  && false  ) { .... }...
\end{lstlisting}

To take measurements, the sensing step, which includes a plan for belief update, as shown in line 1, involves actions such as \emph{sampleColor}. When the sensing action is triggered, it starts the timer until a plan or statement condition is satisfied and \emph{saveResult} action is called. To satisfy all the conditions, each condition of the nested side was bypassed by the \emph{$|true$} expression. In this way, all the satisfied plan conditions/expressions' execution times were recorded. This method was followed for the remaining scenarios and languages. Our aim here is to measure the language's performance based on similar structures. During the measurements, all background applications were stopped, including both PC and RaspberryPi 3.


\subsection{Experimental Evaluation Results}\label{ExperimResults}

In this subsection, experimental evaluation of WCET measurements for six languages is given. As mentioned, both recursive and flat plan/condition/statements were measured starting from perceiving data to acting on each call. At first, overall experimental analyses including both PC and RaspberryPi 3 are given. Then, the PC and RaspberryPi 3 analyses are presented comparatively.

\subsubsection{Normal Plan/Condition Calls for overall performance}

\begin{figure}[H]
\centering
\includegraphics[width=0.6\columnwidth]{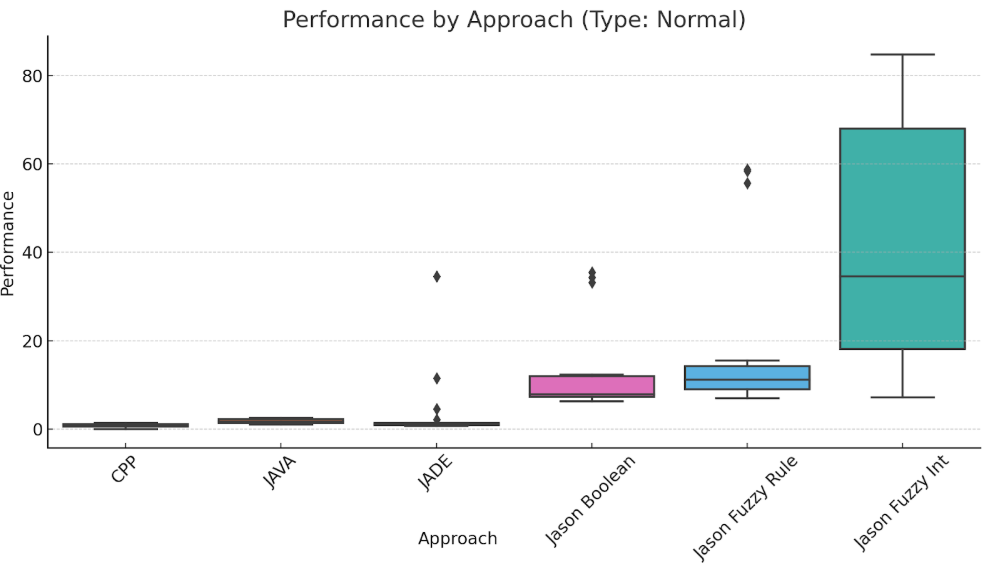}
\caption{Box plot of normal call execution time performance of each selected language based on rule size in milliseconds.}
\label{fig:NormalBoxPlot}
\end{figure}

Figure \ref{fig:NormalBoxPlot} shows a box plot of each language's regular call. Figure shows an overview of each approach’s performance, revealing variations, medians, and outliers. The “Jason Fuzzy Integrated” approach displays the widest range and highest maximum performance, exceeding a value of 80, suggesting strong potential in high-complexity scenarios. However, its extensive interquartile range also reflects performance variability. The “Jason Fuzzy Rule” and “Jason Boolean” approaches exhibit more compact distributions with moderate median values, indicating more predictable and stable execution times. Jade presents a few performance spikes as outliers, but its general distribution remains low. Similarly, Java and C++ approaches have the lowest medians and narrow performance ranges, indicating consistent but subpar results. Overall, the boxplot reinforces the observation that fuzzy-based Jason approaches are substantially more costly in handling normal plan/condition writing scenarios, especially as plan complexity increases.

\begin{figure}[H]
\centering
\includegraphics[width=0.6\columnwidth]{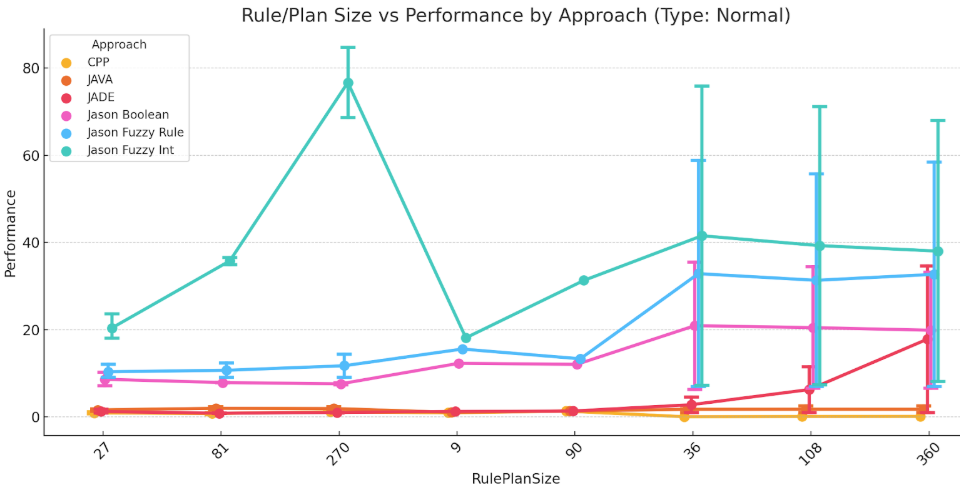}
\caption{Line plot of normal call execution time performance of each selected language based on rule size in milliseconds.}
\label{fig:LineBoxPlot}
\end{figure}

Figure \ref{fig:LineBoxPlot} illustrates the relationship between rule/plan size and the performance of each approach, including error bars that represent performance variance. Once again, the “Jason Fuzzy Int” approach dominates the graph, particularly at the rule size of 270, where performance reaches its highest point. However, this approach also exhibits noticeable variance, which suggests that while it can achieve superior outcomes, it may also produce less stable results under certain conditions. In contrast, both the “Jason Fuzzy Rule” and “Jason Boolean” approaches maintain relatively steady performance across different sizes, indicating their robustness and reliability. Meanwhile, the traditional approaches (C++, Java, and Jade) remain clustered at the lower end of the performance scale, showing minimal responsiveness to increased plan/condition/rule complexity.

\begin{figure}[H]
\centering
\includegraphics[width=0.6\columnwidth]{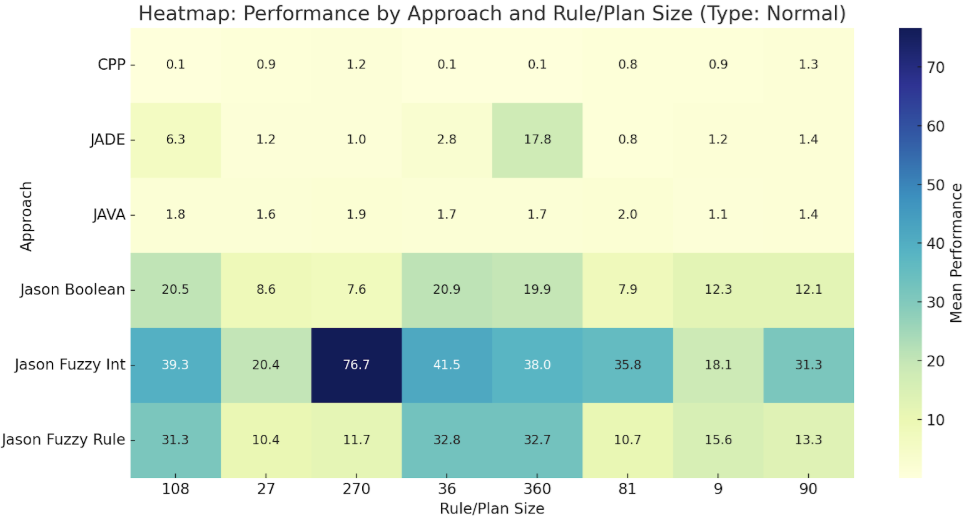}
\caption{Heat map of normal call execution time performance of each selected language based on rule size in milliseconds.}
\label{fig:HeatMapPlot}
\end{figure}

Figure \ref{fig:HeatMapPlot} presents the heat map of six languages across varying rule/plan sizes. A striking observation is the significant performance of the “Jason Fuzzy Int” approach, which peaks at 76.7 when the rule/plan size reaches 270. This approach consistently outperforms the others across most sizes, highlighting its strong adaptability and effectiveness in complex planning environments. The “Jason Fuzzy Rule” approach also delivers relatively high and consistent performance, though not as exceptional as its fuzzy-integrated counterpart. On the other hand, the “Jason Boolean” method yields moderate performance levels and appears stable but less impressive. Traditional approaches such as C++, Java, and Jade demonstrate significantly lower performance across all plan/condition/rule sizes, indicating their limited scalability and efficiency in handling complex agent behaviour under normal conditions.

Overall, the Jason Fuzzy Integrated approach demonstrates the highest WCET. While it can deliver exceptional outcomes, its results may fluctuate depending on the scenario. In contrast, “Jason Fuzzy Rule” provides a more balanced trade-off between performance and consistency, making it a reliable choice for systems where moderate to high performance is required with predictable behaviour. The “Jason Boolean” method, while not as powerful as the fuzzy counterparts, maintains stable results and can be sufficient for simpler decision-making environments. On the lower end of the spectrum, “Jade” occasionally shows performance spikes but generally remains underwhelming. Finally, traditional programming languages like Java and C++ consistently deliver the lowest WCET with limited variation, suggesting that while they may be suitable for basic or legacy systems, they are not well-suited for handling adaptive or large-scale agent-based reasoning tasks. However, on the non-multiplied versions of the case studies, which are 9, 27, and 36 for Network Scaler, Cleaning Robots, and Smart Production Line, it performs considerably better in terms of WCET.

\begin{table}[ht]
\centering
\caption{ANOVA results showing the influence of factors on performance}
\label{tab:anova_resultsNormal}
\begin{tabular}{lccc}
\toprule
\textbf{Factor} & \textbf{F-value} & \textbf{p-value} & \textbf{Significant?} \\
\midrule
Language & 27.69 & $<$ 0.0001 &  Yes \\
RulePlanSize & 2.21 & 0.042 &  Yes \\
Language × RulePlanSize & 17533.00 & 0.077 & No (borderline) \\
\bottomrule
\end{tabular}
\end{table}

Table \ref{tab:anova_resultsNormal} shows that the ANOVA results reveal that the Language factor has a statistically significant effect on system behaviour (F = 27.69, p $<$ 0.0001), indicating that the choice of method substantially influences performance outcomes. Additionally, the RulePlanSize factor demonstrates a marginal yet statistically significant effect (p = 0.042), suggesting that the size of rule-based plans also contributes to variation in system behaviour, albeit to a lesser extent. However, the interaction between Language and RulePlanSize was not statistically significant (p = 0.077). While this value is close to the conventional threshold of significance, it does not provide sufficient evidence to confirm a meaningful interaction effect between these two factors.

\begin{table}[ht]
\centering
\caption{Descriptive statistics of execution time across different approaches}
\label{tab:descriptive_statsNormal}
\begin{tabular}{lccccc}
\toprule
\textbf{Language} & \textbf{Mean} & \textbf{Std Dev} & \textbf{Min} & \textbf{Median} & \textbf{Max} \\
\midrule
C++ & 0.83 & 0.43 & 0.05 & 1.03 & 1.33 \\
Jade & 3.33 & 7.55 & 0.73 & 1.03 & 34.60 \\
Java & 1.74 & 0.54 & 1.00 & 1.67 & 2.49 \\
Jason Boolean & 11.99 & 9.52 & 6.32 & 7.89 & 35.51 \\
Jason Fuzzy Int & 39.94 & 26.87 & 7.21 & 34.60 & 84.78 \\
Jason Fuzzy Rule & 17.33 & 17.06 & 6.92 & 11.13 & 58.76 \\
\bottomrule
\end{tabular}
\end{table}

Table \ref{tab:descriptive_statsNormal} shows descriptive statistics that provide insight into the variability and performance characteristics of each method. C++ exhibits the lowest mean (0.83) and standard deviation (0.43), indicating that it is the fastest and most consistent approach. Jade, on the other hand, shows considerable variability, with a high standard deviation (7.55) and a maximum value that deviates significantly from the median, suggesting unstable performance across trials. Java demonstrates moderate mean execution times (1.74) and relatively low variability, indicating a balanced and stable behaviour. Jason Boolean shows higher mean values (11.99), representing a relatively slower method with moderate variability. Jason Fuzzy Integrated stands out with the highest mean (39.94) and the highest standard deviation (26.87), indicating its high computational cost and inconsistency. Jason Fuzzy Rule, while also slower, performs more stably compared to the Fuzzy Int variant, as reflected in its lower standard deviation (17.06).

\begin{table}[ht]
\centering
\caption{Pairwise comparisons between approaches (Post-hoc test)}
\label{tab:pairwise_comparisonsNormal}
\begin{tabular}{lllcccc}
\toprule
\textbf{Group 1} & \textbf{Group 2} & \textbf{Mean Diff} & \textbf{p-adj} & \textbf{Lower} & \textbf{Upper} & \textbf{Reject} \\
\midrule
C++ & Jade & 2.50 & 0.2261 & -0.71 & 5.71 & False \\
C++ & Java & 0.91 & 0.9698 & -2.20 & 4.03 & False \\
C++ & Jason Boolean & 11.16 & 0.0010 & 7.94 & 14.37 & True \\
C++ & Jason Fuzzy Int & 39.11 & 0.0010 & 35.89 & 42.32 & True \\
C++ & Jason Fuzzy Rule & 16.51 & 0.0010 & 13.30 & 19.73 & True \\
Jade & Java & -1.59 & 0.7507 & -4.62 & 1.44 & False \\
Jade & Jason Boolean & 8.66 & 0.0010 & 5.63 & 11.70 & True \\
Jade & Jason Fuzzy Int & 36.61 & 0.0010 & 33.58 & 39.64 & True \\
Jade & Jason Fuzzy Rule & 14.01 & 0.0010 & 10.98 & 17.04 & True \\
Java & Jason Boolean & 10.25 & 0.0010 & 7.22 & 13.28 & True \\
Java & Jason Fuzzy Int & 38.20 & 0.0010 & 35.17 & 41.23 & True \\
Java & Jason Fuzzy Rule & 15.60 & 0.0010 & 12.57 & 18.63 & True \\
Jason Boolean & Jason Fuzzy Int & 27.95 & 0.0010 & 24.92 & 30.98 & True \\
Jason Boolean & Jason Fuzzy Rule & 5.35 & 0.0010 & 2.32 & 8.38 & True \\
Jason Fuzzy Int & Jason Fuzzy Rule & -22.60 & 0.0010 & -25.63 & -19.57 & True \\
\bottomrule
\end{tabular}
\end{table}

Table \ref{tab:pairwise_comparisonsNormal} shows that the post-hoc pairwise comparisons reveal statistically significant differences between several of the tested methods. No significant differences were found between C++, Jade, and Java, suggesting these approaches perform comparably. However, all comparisons involving C++ versus Jason-based approaches (Boolean, Fuzzy Integrated, Fuzzy Rule) yielded significant differences (p $<$ 0.001), with Jason-based approaches performing significantly slower. Similar patterns were observed when comparing Jade and Java to Jason-based methods, further confirming the substantial performance gap. Within the Jason-based group, Jason Fuzzy Int was found to be significantly slower than both Jason Boolean and Jason Fuzzy Rule, with the difference between Fuzzy Integrated and Fuzzy Rule also being statistically significant. These findings suggest that while fuzzy reasoning provides richer expressiveness, it incurs a measurable computational cost, particularly for Jason Fuzzy Integrated, which consistently shows the highest overhead.

\subsubsection{Plan/Condition Recursive Calls for overall performance}

\begin{figure}[H]
\centering
\includegraphics[width=0.6\columnwidth]{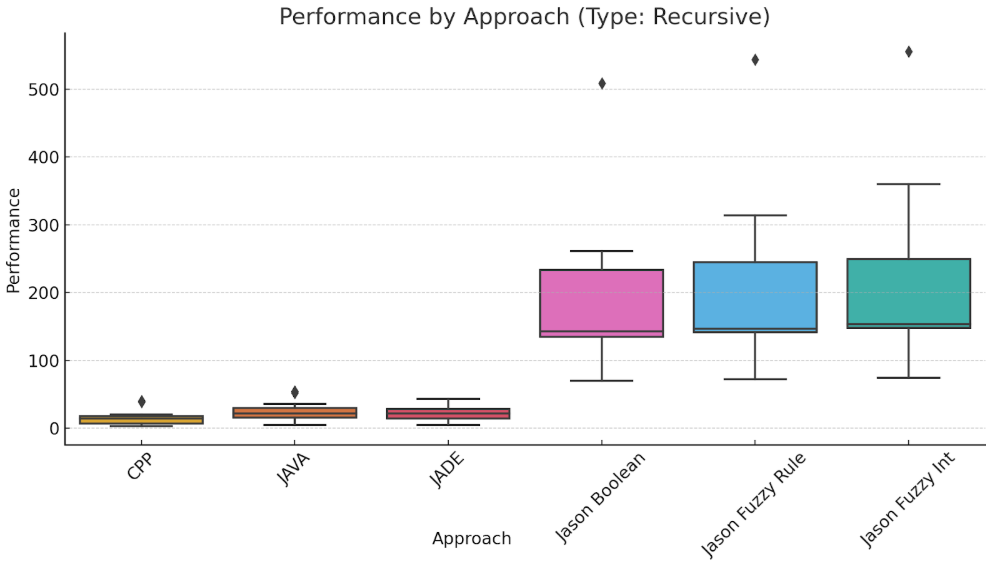}
\caption{Box plot of recursive call execution time performance of each selected language based on rule size in milliseconds.}
\label{fig:RecursiveBoxPlot}
\end{figure}

Figure \ref{fig:RecursiveBoxPlot} visualizes the distribution of WCET values for each approach. C++, Java, and Jade show compact and low distributions, reflecting consistent and efficient performance. On the other hand, Jason-based approaches have higher average WCETs, wider spreads (indicating higher variance), and several outliers. These outliers suggest that in some scenarios, the Jason approaches experience extreme delays, making them less reliable for time-critical applications.

\begin{figure}[H]
\centering
\includegraphics[width=0.6\columnwidth]{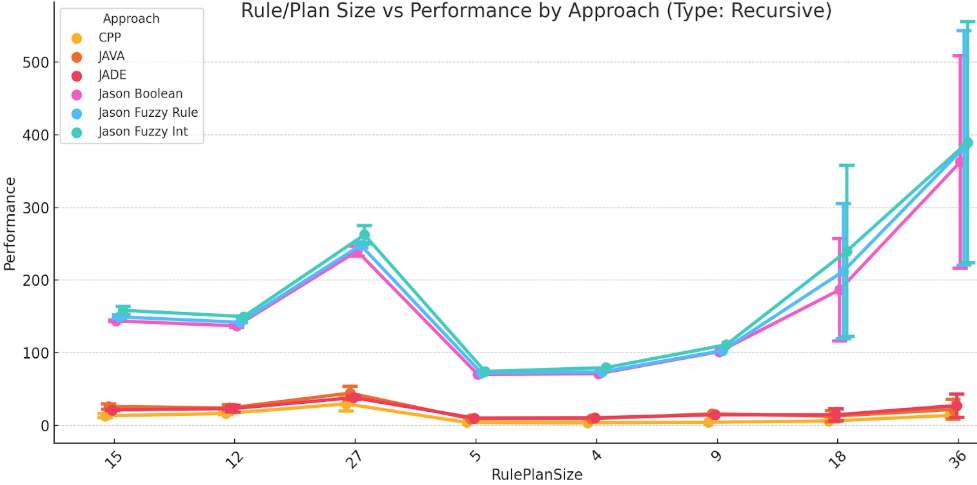}
\caption{Line plot of recursive call execution time performance of each selected language based on rule size in milliseconds.}
\label{fig:RecursiveLinePlot}
\end{figure}

Figure \ref{fig:RecursiveLinePlot}  shows the relationship between plan size (X-axis) and WCET performance (Y-axis). Jason-based methods (represented in blue, pink, and cyan) show a steep rise in WCET as the plan size increases, especially at sizes 27 and 36, where performance gaps become significant. In contrast, C++, Java, and Jade remain stable with low WCET values across all plan sizes. Additionally, the error bars for Jason's approaches are large, indicating high variability and unstable behaviour across runs.

\begin{figure}[H]
\centering
\includegraphics[width=0.6\columnwidth]{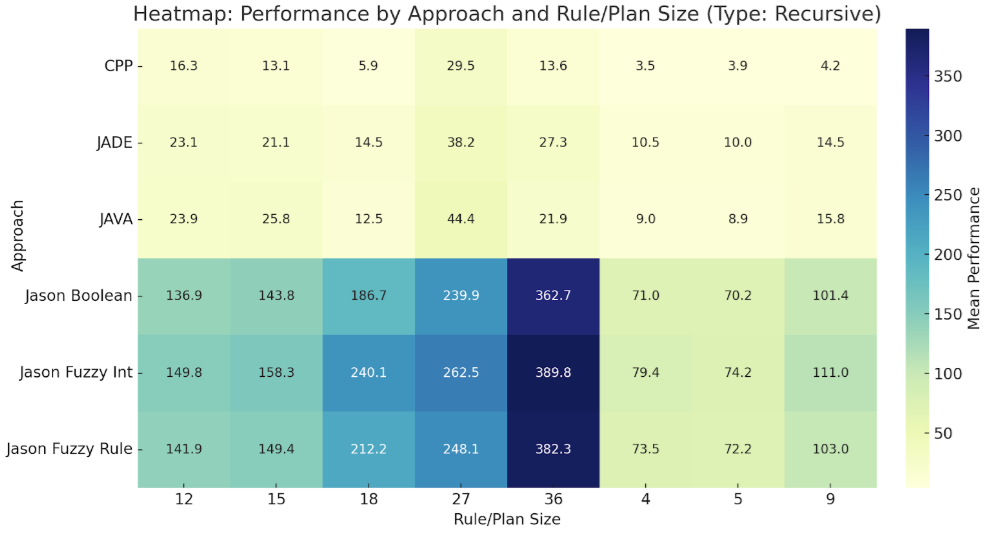}
\caption{Heat map of normal call execution time performance of each selected language based on rule size in milliseconds.}
\label{fig:RecursiveHeatMapPlot}
\end{figure}

Figure \ref{fig:RecursiveHeatMapPlot} 
illustrates the \textbf{average} WCET across different plan/condition/rule sizes for each approach. C++, Jade, and Java consistently achieve lower WCET values, indicating better performance. In contrast, Jason-based approaches (Boolean, Fuzzy Int, and Fuzzy Rule) show significantly higher WCETs, especially as the plan size increases. At a plan size of 36, Jason Fuzzy Int reaches 389.8, Fuzzy Rule 382.3, and Boolean 362.7, while C++ remains as low as 13.6. This highlights a clear performance degradation in Jason approaches with larger recursive structures, even when smaller plan sizes are used. The best performance (i.e., lowest WCET) is observed in the C++ implementation, followed closely by Java and Jade. The high performance is seen in Jason-based methods in general, particularly Fuzzy Integrated and Fuzzy Rule, followed by Jason Boolean. As plan/condition/rule complexity increases, Jason approaches start to demand more hardware performance and time.

Ultimately, the increase in recursive and normal scenarios demonstrates that the average WCET performance improves significantly for all approaches, indicating the added complexity and cost of handling recursion. The choice of approach has a meaningful impact on performance in both recursive and non-recursive contexts. However, the influence of rule, condition or plan size is significant in recursive scenarios, where larger structures result in a steep rise in execution time. Jason-based methods, particularly those using fuzzy logic, consistently result in higher computational cost. Among these, Jason Fuzzy Integrated emerges as the most expensive approach in terms of WCET. In summary, while Jason-based systems offer flexible agent reasoning, their performance in recursive and large-scale plans is an explicit limitation. However, such plan/condition/rule requirements are relatively futuristic or too specific for case studies and may also indicate poor fuzzy function design.

\begin{table}[ht]
\centering
\caption{Results of Two-Way ANOVA for Language and Rule Plan Size}
\label{tab:anova_resultsRec}
\begin{tabular}{lccc}
\hline
\textbf{Factor} & \textbf{F-value} & \textbf{p-value} & \textbf{Significant?} \\
\hline
Language & 50.38 & $<$0.0001 & Yes \\
Rule Plan Size & 9.76 & $<$0.0001 & Yes \\
Language $\times$ Rule Plan Size & 1.45 & 0.089 & No (Marginal) \\
\hline
\end{tabular}
\end{table}

As presented in Table~\ref{tab:anova_resultsRec}, the ANOVA results indicate that both the Language and RulePlanSize factors have statistically significant effects on the dependent variable (F = 50.38, p $<$ 0.0001 and F = 9.76, p $<$ 0.0001, respectively). This suggests that the choice of method, as well as the complexity of the rule-based plan, significantly influences system performance. However, the interaction term (Language × RulePlanSize) is not statistically significant (p = 0.089), although it remains close to the conventional significance threshold. This marginal result suggests that while both factors have an individual impact, their combined influence does not produce a synergistic or interfering effect on the outcome variable.

\begin{table}[ht]
\centering
\caption{Descriptive statistics by approach}
\label{tab:descriptive_statisticsRec}
\begin{tabular}{lccccc}
\hline
\textbf{Language} & \textbf{Mean} & \textbf{Std Dev} & \textbf{Min} & \textbf{Median} & \textbf{Max} \\
\hline
C++ & 15.17 & 10.49 & 3.51 & 14.39 & 40.20 \\
Jade & 22.70 & 11.36 & 4.93 & 21.98 & 43.37 \\
Java & 23.99 & 13.75 & 4.41 & 21.76 & 54.44 \\
Jason Boolean & 180.82 & 96.65 & 70.15 & 142.72 & 509.02 \\
Jason Fuzzy Int & 204.14 & 114.07 & 74.18 & 153.12 & 555.76 \\
Jason Fuzzy Rule & 191.40 & 106.64 & 72.18 & 146.83 & 543.85 \\
\hline
\end{tabular}
\end{table}

The descriptive statistics in Table~\ref{tab:descriptive_statisticsRec} further elucidate the performance characteristics of each approach. C++ yields the lowest mean execution time (15.17) with moderate variability, indicating relatively efficient and consistent behaviour. Both Jade and Java exhibit slightly higher mean values (22.70 and 23.99, respectively), but still within the same order of magnitude, suggesting similar mid-level performance with tolerable variance. In contrast, the Jason-based approaches (Jason Boolean, Jason Fuzzy Int, and Jason Fuzzy Rule) show dramatically higher means (ranging from 180.82 to 204.14) and standard deviations, revealing that they are significantly more computationally intensive and less stable. These findings highlight a clear divide between lightweight and cognitively rich agent platforms in terms of execution efficiency.

\begin{table}[htbp]
\centering
\caption{Multiple Comparison Test Results Between Groups}
\label{tab:group_comparisonRec}
\begin{tabular}{l l r r r r l}
\hline
\textbf{group1} & \textbf{group2} & \textbf{meandiff} & \textbf{p-adj} & \textbf{lower} & \textbf{upper} & \textbf{reject} \\
\hline
C++             & Jade            & 7.53    & 0.0179 & 0.56   & 14.49  & True  \\
C++             & Java            & 8.83    & 0.0034 & 1.86   & 15.80  & True  \\
C++             & Jason Boolean   & 165.66  & 0.0010 & 158.69 & 172.62 & True  \\
C++             & Jason Fuzzy Int & 188.97  & 0.0010 & 182.00 & 195.93 & True  \\
C++             & Jason Fuzzy Rule& 176.23  & 0.0010 & 169.26 & 183.19 & True  \\
Jade            & Java            & 1.30    & 0.9982 & -5.67  & 8.27   & False \\
Jade            & Jason Boolean   & 158.13  & 0.0010 & 151.16 & 165.10 & True  \\
Jade            & Jason Fuzzy Int & 181.44  & 0.0010 & 174.47 & 188.41 & True  \\
Jade            & Jason Fuzzy Rule& 168.70  & 0.0010 & 161.73 & 175.67 & True  \\
Java            & Jason Boolean   & 156.83  & 0.0010 & 149.86 & 163.80 & True  \\
Java            & Jason Fuzzy Int & 180.13  & 0.0010 & 173.16 & 187.10 & True  \\
Java            & Jason Fuzzy Rule& 167.40  & 0.0010 & 160.43 & 174.37 & True  \\
Jason Boolean   & Jason Fuzzy Int & 23.31   & 0.0030 & 16.34  & 30.28  & True  \\
Jason Boolean   & Jason Fuzzy Rule& 10.57   & 0.2876 & 3.60   & 17.54  & False \\
Jason Fuzzy Int & Jason Fuzzy Rule& -12.74  & 0.1253 & -19.71 & -5.77  & False \\
\hline
\end{tabular}
\end{table}

According to the post-hoc pairwise comparisons in Table~\ref{tab:group_comparisonRec}, there are several statistically significant performance differences between the evaluated methods. Specifically, C++ is significantly faster than all other approaches except Jade and Java, confirming its superior efficiency. Likewise, both Jade and Java outperform the Jason-based approaches with statistically significant differences. No significant difference is observed between Jade and Java, indicating their performance is statistically comparable. Among the Jason-based methods, Jason Fuzzy Int is significantly slower than Jason Boolean (p = 0.003), while Jason Fuzzy Rule does not differ significantly from either. The comparison between Jason Fuzzy Int and Jason Fuzzy Rule yields a non-significant result (p = 0.1253), although the mean difference is noticeable. These results collectively reinforce that while fuzzy and rule-based reasoning may enhance decision quality, they come at a substantial computational cost, particularly in the case of fuzzy intention-based control.

Overall, Post-hoc comparisons further confirm that Jason-based approaches are significantly slower than their lightweight counterparts, with Jason Fuzzy Integrated emerging as the most computationally intensive. Interestingly, although fuzzy logic enhances the richness of decision-making, its implementation in intention-based reasoning introduces non-trivial overhead. These findings suggest that in real-time or resource-constrained environments, simpler platforms may provide a more favourable trade-off between performance and reasoning complexity. Therefore, selecting an agent architecture should be carefully aligned with system-level requirements for speed, consistency, and cognitive load.

\subsubsection{PC and RaspberryPi 3 Comparison}

This part presents a comparative performance analysis of six programming approaches (C++, Jade, Java, Jason Boolean, Jason Fuzzy Int, and Jason Fuzzy Rule) on both PC and Raspberry PI 3 platforms. The focus is on investigating how method types (Normal vs. Recursive) and increasing rule/plan sizes affect execution times.  Statistical analyses such as ANOVA and Tukey HSD tests complement the visual insights, confirming the observed trends. Overall, this section emphasizes the trade-offs between computational efficiency and the advanced reasoning capabilities of Jason-based systems and conventional ones in comparison of PC and RaspberryPi 3. The Smart Production Line (PL) is the most and compatible case study that can be run on RaspberryPi 3. Therefore, the experiments and analyses were carried out accordingly.

\begin{table}[htbp]
\centering
\caption{Execution Time Statistics of Six Languages on PC and RP3}
\label{tab:lang_performancePCvsRP3}
\begin{tabular}{l r r r r r r r}
\hline
\textbf{Language} & \textbf{count} & \textbf{mean} & \textbf{std} & \textbf{min} & \textbf{25\%} & \textbf{50\%} & \textbf{max} \\
\hline
C++              & 36.0 & 7.99  & 10.31  & 0.05 & 1.06 & 2.42  & 40.20 \\
Jade             & 42.0 & 13.01 & 13.67  & 0.73 & 1.04 & 8.74  & 43.37 \\
Java             & 42.0 & 12.86 & 14.81  & 1.00 & 1.69 & 3.45  & 54.44 \\
Jason Boolean    & 42.0 & 96.41 & 109.09 & 6.32 & 7.89 & 52.83 & 509.02 \\
Jason Fuzzy Int  & 42.0 & 122.04 & 116.64 & 7.21 & 34.89 & 82.08 & 555.76 \\
Jason Fuzzy Rule & 42.0 & 104.37 & 115.97 & 6.92 & 11.40 & 65.47 & 543.85 \\
\hline
\end{tabular}
\end{table}

Table~\ref{tab:lang_performancePCvsRP3} provides the mean execution times, standard deviations, and extreme values of different approaches. According to this table, C++ is by far the fastest approach. At the same time, Jason-based methods, particularly Jason Fuzzy Integrated and Jason Fuzzy Rule, exhibit significantly longer execution times and higher variance.

\begin{table}[htbp]
\centering
\caption{Two-Way ANOVA Results for PC vs. RP3 and Method Type}
\label{tab:anova_resultsPCvsRP3}
\begin{tabular}{l c c c}
\hline
\textbf{Metric} & \textbf{Case Effect} & \textbf{Type Effect} & \textbf{Interaction} \\
\hline
Java            & Significant (p = 0.0047)  & Significant (p = 0.0008)  & Significant (p = 0.0106) \\
Jason Boolean   & Significant (p = 0.0418)  & Significant (p = 0.0011)  & Not Significant (p = 0.1087) \\
Jason Fuzzy Rule& Significant (p = 0.0114)  & Significant (p = 0.0006)  & Marginal (p = 0.0696) \\
Jason Fuzzy Int & Significant (p = 0.0021)  & Significant (p = 0.0001)  & Significant (p = 0.0254) \\
\hline
\end{tabular}
\end{table}

The two-way ANOVA analysis shown by Table~\ref{tab:anova_resultsPCvsRP3} reveals that both the hardware platform (PC vs RP3) and the method type (Normal vs Recursive) have significant effects on performance. In particular, Recursive methods run much slower on RP3, and this difference is statistically significant. Additionally, for certain metrics (e.g., Java and Jason Fuzzy Int), the interaction between platform and method type is significant, meaning that the impact of the Recursive method depends on the platform used.

\begin{table}[htbp]
\centering
\caption{Descriptive Statistics for PC vs RP3 based on Jason Boolean}
\label{tab:descriptive_statisticsPCvsRP3}
\begin{tabular}{l l r r r}
\hline
\textbf{Case} & \textbf{Type} & \textbf{Mean} & \textbf{Std Dev} & \textbf{N} \\
\hline
PL RP3 & Normal    & 34.34   & 1.19   & 3 \\
PL RP3 & Recursive & 340.93  & 145.64 & 3 \\
PL PC   & Normal    & 6.49    & 0.15   & 3 \\
PL PC   & Recursive & 149.80  & 57.73  & 3 \\
\hline
\end{tabular}
\end{table}

When Table~\ref{tab:descriptive_statisticsPCvsRP3} is examined, it is observed that the Jason Boolean approach exhibits a very high average execution time (340.93 ms) on the RP3 platform in the Recursive type, which is statistically significantly higher than both the Normal type and the values on the PC platform. While the lowest average (6.49 ms) is recorded for the Normal type on the PC, the RP3 Recursive configuration has the highest variance and standard deviation. These findings indicate that Recursive methods cause significant performance bottlenecks on embedded hardware such as RP3. Specifically, the Smart Production Line case study (PL) executed on RP3 with Recursive calls has the highest execution time compared to all other cases. The Tukey HSD post-hoc analysis (Table~\ref{tab:tukey_resultsPCvsRP3}) further confirms that these differences are statistically significant.

\begin{table}[htbp]
\centering
\caption{Significant Pairwise Differences (Tukey HSD)}
\label{tab:tukey_resultsPCvsRP3}
\begin{tabular}{l l c}
\hline
\textbf{Metric} & \textbf{Comparison} & \textbf{p-value} \\
\hline
Java            & PL RP3-Normal vs PL RP3-Recursive & 0.0014 \\
                & PL RP3-Recursive vs PL PC-Normal   & 0.0009 \\
                & PL RP3-Recursive vs PL PC-Recursive& 0.0042 \\
Jason Boolean   & PL RP3-Normal vs PL RP3-Recursive & 0.0060 \\
                & PL RP3-Recursive vs PL PC-Normal   & 0.0035 \\
Jason Fuzzy Rule& PL RP3-Normal vs PL RP3-Recursive & 0.0031 \\
                & PL RP3-Recursive vs PL PC-Normal   & 0.0012 \\
                & PL RP3-Recursive vs PL PC-Recursive& 0.0221 \\
Jason Fuzzy Int & PL RP3-Normal vs PL RP3-Recursive & 0.0007 \\
                & PL RP3-Recursive vs PL PC-Normal   & 0.0002 \\
                & PL RP3-Recursive vs PL PC-Recursive& 0.0041 \\
\hline
\end{tabular}
\end{table}

\begin{figure}[H]
\centering
\includegraphics[width=0.65\columnwidth]{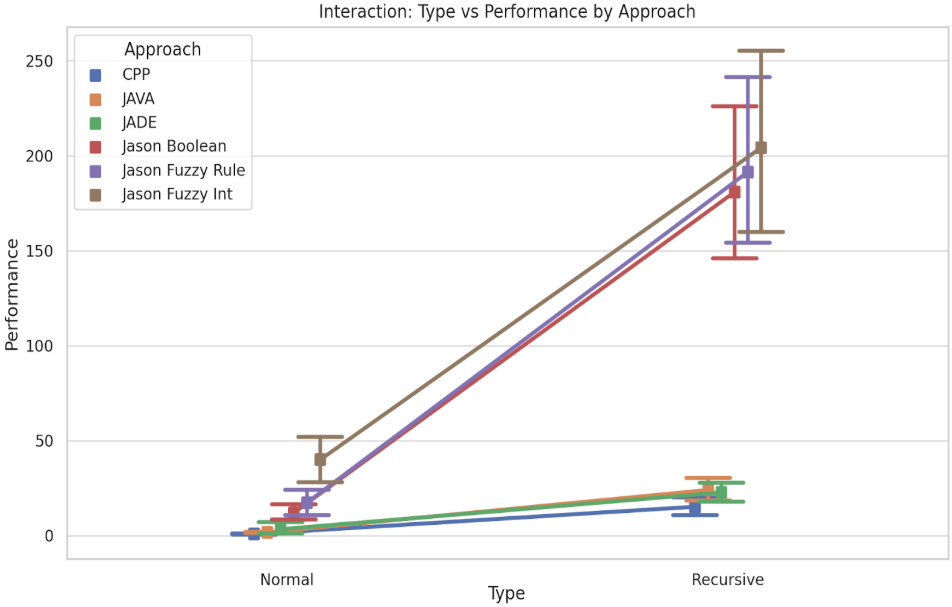}
\caption{Comparison of Normal vs Recursive performance for each approach.}
\label{fig:TypeVsApproachPCvsRP3}
\end{figure}

\begin{figure}[H]
\centering
\includegraphics[width=0.65\columnwidth]{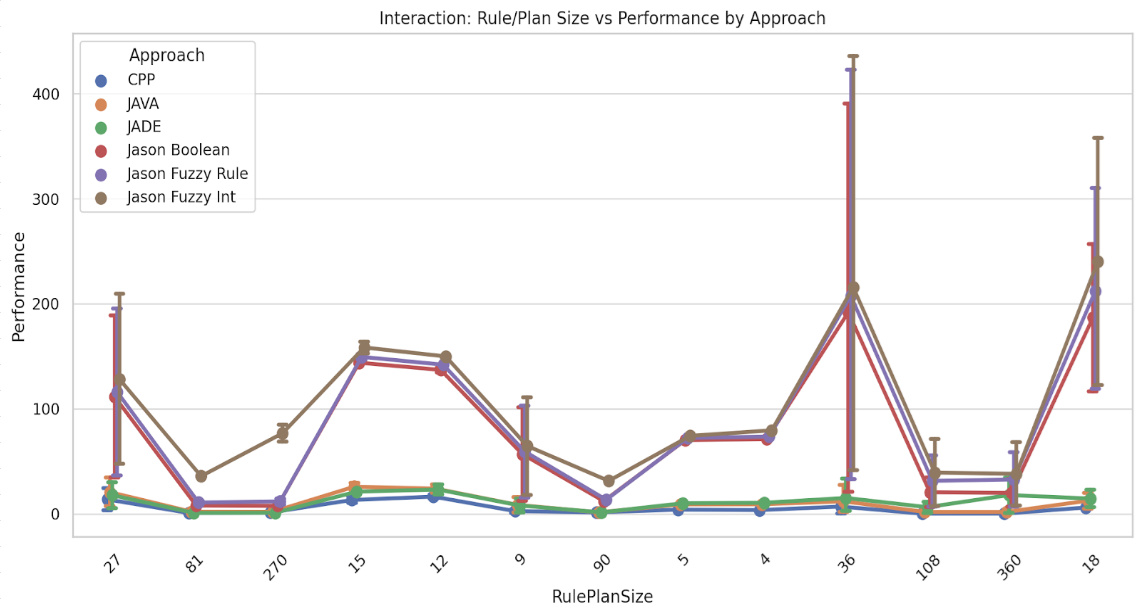}
\caption{Line plot showing the impact of rule and plan size on execution performance across approaches.}
\label{fig:RulePlanLineChartPCvsRP3}
\end{figure}

\begin{figure}[H]
\centering
\includegraphics[width=0.65\columnwidth]{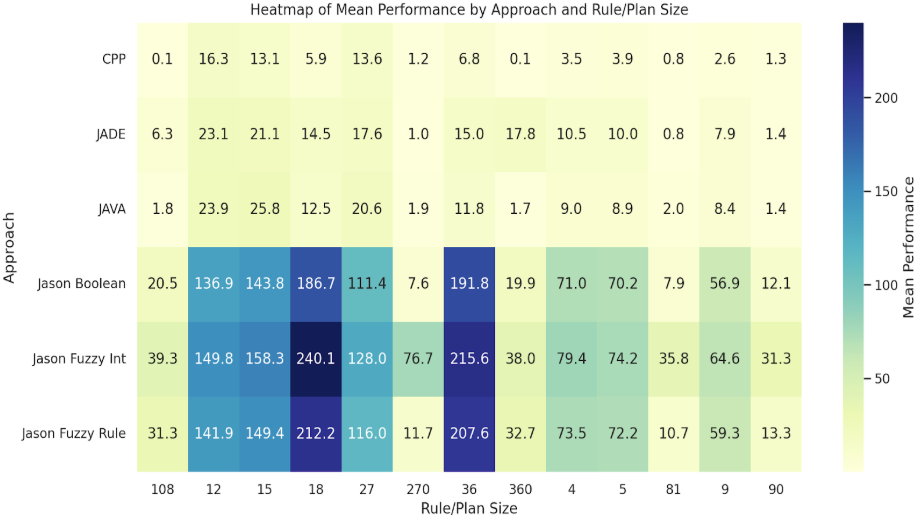}
\caption{Heatmap visualisation of average execution times highlighting performance variations by approach and configuration.}
\label{fig:HeatmapPerformancePCvsRP3}
\end{figure}

\begin{figure}[H]
\centering
\includegraphics[width=0.65\columnwidth]{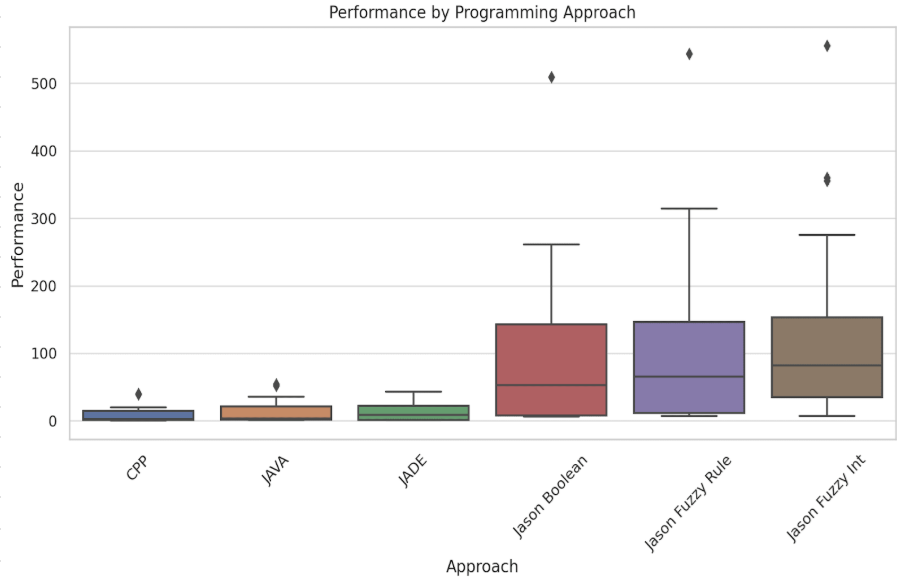}
\caption{BarPlot visualisation of average execution times highlighting performance variations by approach and configuration.}
\label{fig:BarplotPerformanceApproach}
\end{figure}

Figure~\ref{fig:TypeVsApproachPCvsRP3} indicates a clear performance gap between Normal and Recursive types, with Recursive methods causing substantial delays, especially for Jason-based systems. Figure~\ref{fig:RulePlanLineChartPCvsRP3} demonstrates that as rule/plan size increases, Jason methods degrade significantly. Figure~\ref{fig:HeatmapPerformancePCvsRP3} visualizes average performance, emphasizing Jason Fuzzy Int’s high time cost under large RulePlanSize scenarios. Additionally, the barplot in Figure~\ref{fig:BarplotPerformanceApproach} further confirms the broader variance and generally higher execution times associated with Jason-based approaches, particularly under fuzzy logic configurations.

Overall, Recursive methods drastically increase execution times, with this effect being particularly pronounced on the RP3 hardware. While the performance of Normal methods on PC and RP3 platforms is relatively similar, RP3 Recursive shows a substantial slowdown compared to all other variants. Although Jason-based approaches enhance cognitive capabilities, they come with high computational costs. However, as shown by  Figure \ref{fig:TypeVsApproachPCvsRP3} such high execution times such as Jason Fuzzy Integrated with 555.76 ms on Table \ref{tab:lang_performancePCvsRP3}) occur when there are 36 recursive calls. This indicates that lightweight approaches such as C++, Jade, or Java may be more suitable for real-time or resource-constrained environments for recursion-intensive applications. In the following subsection, empirical evaluation results are presented.

\subsection{Empirical Evaluation Results}\label{EmpiricalResults}

In this subsection, empirical evaluation results are presented both language- and group-wise.

\subsubsection{Language-wise Results}

Empirical evaluations are presented for the six programming languages and then three groups. Table \ref{tab:EmpiricalPerformanceDevelopment} presents statistics on the total, plan/rule/condition writing, deliberation, and debugging phases of the performed evaluations, based on the selected languages. In addition, this plan/rule/condition phase also covers the requirements introduced in subsections \ref{subsec55}, \ref{subsec56} and \ref{subsec57}.

\begin{table}[H]
\centering
\scriptsize
\resizebox{\textwidth}{!}{%
\begin{tabular}{|l|
                |c|c|c|c|c
                |c|c|c|c|c
                |c|c|c|c|c
                |c|c|c|c|c|}
\hline
\textbf{Name} & 
\multicolumn{5}{c|}{\textbf{Total Time}} & 
\multicolumn{5}{c|}{\textbf{Plan Writing}} & 
\multicolumn{5}{c|}{\textbf{Deliberation}} & 
\multicolumn{5}{c|}{\textbf{Debug}} \\
\hline
 & Mean & Std & Min & Max & Cnt. 
 & Mean & Std & Min & Max & Cnt. 
 & Mean & Std & Min & Max & Cnt. 
 & Mean & Std & Min & Max & Cnt. \\
\hline
Jason Fuzzy Int.  & 25.33 & 7.40 & 11 & 44 & 21 & 13.52 & 4.45 & 8 & 24 & 21 & 4.86 & 1.06 & 3 & 7 & 21 & 2.95 & 1.50 & 1 & 6 & 21 \\
\hline
Jason Rule Fuzzy & 31.62 & 8.63 & 14 & 52 & 21 & 18.62 & 6.12 & 10 & 35 & 21 & 5.43 & 1.54 & 3 & 10 & 21 & 2.67 & 1.24 & 1 & 5 & 21 \\
\hline
Jason Bool & 28.38 & 8.04 & 12 & 39 & 21 & 12.95 & 5.08 & 6 & 23 & 21 & 5.14 & 1.80 & 3 & 8 & 21 & 2.43 & 0.98 & 1 & 4 & 21 \\
\hline
Jade       & 37.19 & 8.41 & 19 & 47 & 21 & 17.29 & 5.11 & 10 & 27 & 21 & 6.24 & 1.87 & 3 & 9 & 21 & 2.71 & 1.31 & 1 & 5 & 21 \\
\hline
Java       & 28.95 & 8.06 & 12 & 39 & 21 & 16.10 & 4.67 & 8 & 24 & 21 & 5.81 & 0.93 & 4 & 7 & 21 & 2.24 & 1.04 & 1 & 5 & 21 \\
\hline
C++        & 28.00 & 7.55 & 15 & 39 & 21 & 15.86 & 4.40 & 10 & 25 & 21 & 5.33 & 1.15 & 3 & 7 & 21 & 2.19 & 1.03 & 1 & 4 & 21 \\
\hline
\end{tabular}
}
\caption{Performance statistics (mean, std, min, max, count) for each language and development phases.}
\label{tab:EmpiricalPerformanceDevelopment}
\end{table}

Table \ref{tab:AnovaResultsApproach} shows the applied ANOVA test performed on the differences in six programming languages elaborated by the participants. 

\begin{table}[ht]
\centering
\small
\caption{ANOVA for Language Differences (significance at p$<$0.05)}
\begin{tabular}{|l|c|c|}
\hline
\textbf{Metric} & \textbf{p-value} & \textbf{Significant?} \\
\hline
Total Time      & 0.0001           &  Yes       \\
Plan Writing    & 0.0024           &  Yes       \\
Deliberation    & 0.037            &  Yes       \\
Debug          & 0.271            &  No        \\
\hline
\end{tabular}
\label{tab:AnovaResultsApproach}
\end{table}

Table \ref{tab:ApproachTotalTimeTukey} shows these approach-wise differences, followed by the Tukey test after carrying out ANOVA. The Tukey HSD results for total development time show that Jade differs significantly from multiple others, particularly Languages Jason Fuzzy Integrated, Jason Boolean, Java, and C++. These significant pairwise differences (p $<$ 0.05) suggest that Jade consistently requires more time than most other implementations. 

\begin{table}[H]
\centering
\small
\caption{Language - Total Time Multiple Comparison of Means (Tukey HSD, FWER=0.05)}
\begin{tabular}{|c|c|r|c|r|r|c|}
\hline
comparision group1 & comparision group2 & \multicolumn{1}{c|}{meandiff} & p-adj & \multicolumn{1}{c|}{lower} & \multicolumn{1}{c|}{upper} & reject \\
\hline
C++ & Java & 6.2857  & 0.1215 & -0.8875  & 13.4589 & False \\
C++ & Jade  & 3.0476  & 0.8211 & -4.1256  & 10.2208 & False \\
C++ & Jason Bool & 11.8571 & 0.0001 & 4.6839   & 19.0303 & True  \\
C++ & Jason Rule Fuzzy  & 3.619   & 0.6894 & -3.5542  & 10.7923 & False \\
C++ & Jason Fuzzy Int. & 2.6667  & 0.8897 & -4.5065  & 9.8399  & False \\
Java & Jade  & -3.2381 & 0.7804 & -10.4113 & 3.9351  & False \\
Java & Jason Bool & 5.5714  & 0.2233 & -1.6018  & 12.7446 & False \\
Java & Jason Rule Fuzzy  & -2.6667 & 0.8897 & -9.8399  & 4.5065  & False \\
Java & Jason Fuzzy Int. & -3.619  & 0.6894 & -10.7923 & 3.5542  & False \\
Jade & Jason Bool & 8.8095  & 0.007  & 1.6363   & 15.9827 & True  \\
Jade & Jason Rule Fuzzy  & 0.5714  & 0.9999 & -6.6018  & 7.7446  & False \\
Jade & Jason Fuzzy Int. & -0.381  & 1.0    & -7.5542  & 6.7923  & False \\
Jason Bool & Jason Rule Fuzzy  & -8.2381 & 0.0145 & -15.4113 & -1.0649 & True  \\
Jason Bool & Jason Fuzzy Int. & -9.1905 & 0.0042 & -16.3637 & -2.0173 & True  \\
Jason Rule Fuzzy  & Jason Fuzzy Int. & -0.9524 & 0.9989 & -8.1256  & 6.2208  & False \\
\hline
\end{tabular}
\label{tab:ApproachTotalTimeTukey}
\end{table}

Table \ref{tab:PlanWritingTukey}  shows the Tukey HSD analysis for the Plan Writing phase, revealing statistically significant differences in time requirements between specific approaches. Notably, Jason Rule Fuzzy required significantly more time than both Jason Fuzzy Integrated (p = 0.0158) and Jason Boolean (p = 0.0048), suggesting that its planning process may involve more complexity or less efficiency. These findings point to a meaningful distinction in cognitive load or 

\begin{table}[ht]
\centering
\small
\caption{Language - Plan Writing Multiple Comparison of Means (Tukey HSD, FWER=0.05)}
\begin{tabular}{|c|c|r|c|r|r|c|}
\hline
comparision group1 & comparision group2 & \multicolumn{1}{c|}{meandiff} & p-adj & \multicolumn{1}{c|}{lower} & \multicolumn{1}{c|}{upper} & reject \\
\hline
C++ & Java & 5.0952  & 0.0158  & 0.6218  & 9.5687  & True  \\
C++ & Jade & -0.5714 & 0.9991  & -5.0448 & 3.902   & False \\
C++ & Jason Bool & 3.7619  & 0.1525  & -0.7115 & 8.2353  & False \\
C++ & Jason Rule Fuzzy  & 2.5714  & 0.5576  & -1.902  & 7.0448  & False \\
C++ & Jason Fuzzy Int. & 2.3333  & 0.6581  & -2.1401 & 6.8067  & False \\
Java & Jade & -5.6667 & 0.0048  & -10.1401 & -1.1933 & True  \\
Java & Jason Bool & -1.3333 & 0.9544  & -5.8067 & 3.1401  & False \\
Java & Jason Rule Fuzzy  & -2.5238 & 0.5778  & -6.9972 & 1.9496  & False \\
Java & Jason Fuzzy Int. & -2.7619 & 0.4772  & -7.2353 & 1.7115  & False \\
Jade & Jason Bool & 4.3333  & 0.0634  & -0.1401 & 8.8067  & False \\
Jade & Jason Rule Fuzzy  & 3.1429  & 0.3291  & -1.3306 & 7.6163  & False \\
Jade & Jason Fuzzy Int. & 2.9048  & 0.4191  & -1.5687 & 7.3782  & False \\
Jason Bool & Jason Rule Fuzzy  & -1.1905 & 0.9719  & -5.6639 & 3.2829  & False \\
Jason Bool & Jason Fuzzy Int. & -1.4286 & 0.9394  & -5.902  & 3.0448  & False \\
Jason Rule Fuzzy  & Jason Fuzzy Int. & -0.2381 & 1.0     & -4.7115 & 4.2353  & False \\
\hline
\end{tabular}
\label{tab:PlanWritingTukey}
\end{table}

Table \ref{tab:DeliberationTukey}, the deliberation phase results show far fewer significant differences. Among all pairwise comparisons, only the difference between Jason Fuzzy Integrated and Jade reached statistical significance (p = 0.0274), indicating that Jade demands more time for reasoning and decision-making processes. 

\begin{table}[h]
\centering
\small
\caption{Language - Deliberation Multiple Comparison of Means (Tukey HSD, FWER=0.05)}
\begin{tabular}{|c|c|r|c|r|r|c|}
\hline
comparision group1 &  comparision group2 & \multicolumn{1}{c|}{meandiff} & p-adj & \multicolumn{1}{c|}{lower} & \multicolumn{1}{c|}{upper} & reject \\
\hline
C++ & Java & 0.5714  & 0.7910  & -0.7135 & 1.8564  & False \\
C++ & Jade & 0.2857  & 0.9874  & -0.9992 & 1.5707  & False \\
C++ & Jason Bool & 1.3810  & 0.0274  & 0.0960  & 2.6659  & True  \\
C++ & Jason Rule Fuzzy  & 0.9524  & 0.2709  & -0.3326 & 2.2373  & False \\
C++ & Jason Fuzzy Int. & 0.4762  & 0.8910  & -0.8088 & 1.7611  & False \\
Java & Jade & -0.2857 & 0.9874  & -1.5707 & 0.9992  & False \\
Java & Jason Bool & 0.8095  & 0.4540  & -0.4754 & 2.0945  & False \\
Java & Jason Rule Fuzzy  & 0.3810  & 0.9554  & -0.9040 & 1.6659  & False \\
Java & Jason Fuzzy Int. & -0.0952 & 0.9999  & -1.3802 & 1.1897  & False \\
Jade & Jason Bool & 1.0952  & 0.1419  & -0.1897 & 2.3802  & False \\
Jade & Jason Rule Fuzzy  & 0.6667  & 0.6632  & -0.6183 & 1.9516  & False \\
Jade & Jason Fuzzy Int. & 0.1905  & 0.9981  & -1.0945 & 1.4754  & False \\
Jason Bool & Jason Rule Fuzzy  & -0.4286 & 0.9277  & -1.7135 & 0.8564  & False \\
Jason Bool & Jason Fuzzy Int. & -0.9048 & 0.3266  & -2.1897 & 0.3802  & False \\
Jason Rule Fuzzy  & Jason Fuzzy Int. & -0.4762 & 0.8910  & -1.7611 & 0.8088  & False \\
\hline
\end{tabular}
\label{tab:DeliberationTukey}
\end{table}

\begin{figure}[H]
\centering
\includegraphics[width=0.6\columnwidth]{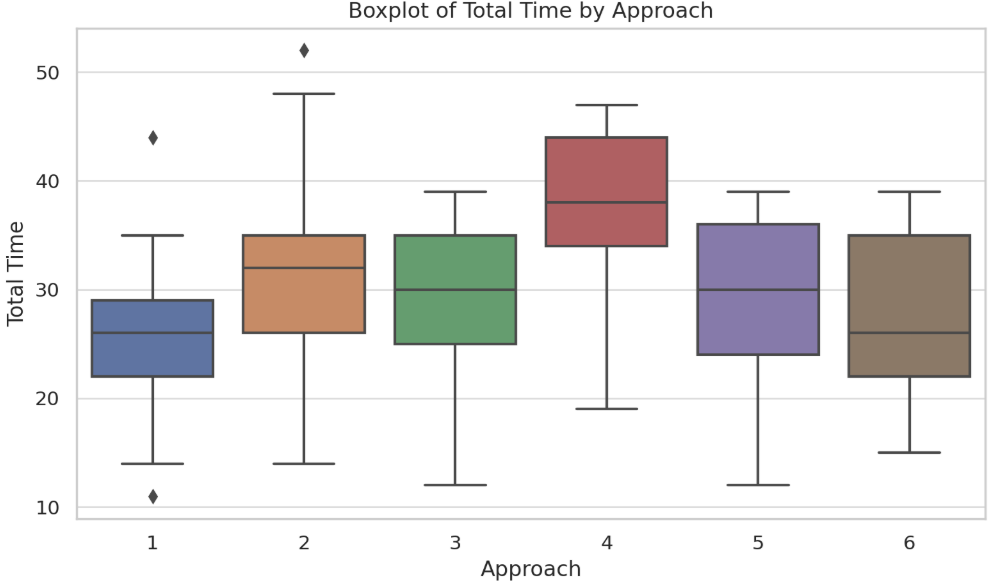}
\caption{BoxPlot Total Development Time based on six selected languages.}
\label{fig:boxPlotByTotalTime}
\end{figure}

\begin{figure}[H]
\centering
\includegraphics[width=0.6\columnwidth]{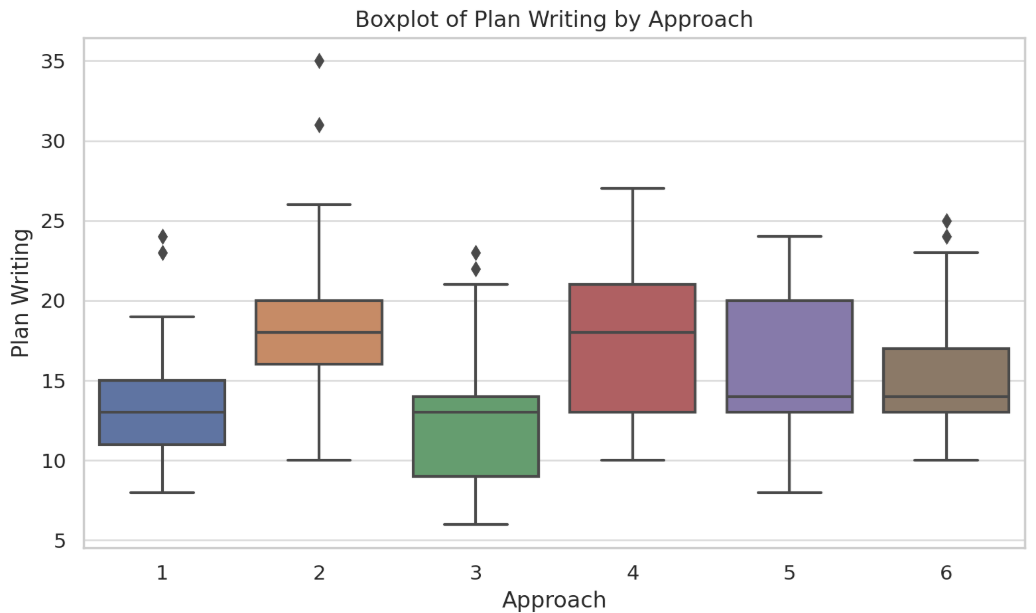}
\caption{BoxPlot Plan Writing Times based on six selected languages.}
\label{fig:boxPlotByPlanWriting}
\end{figure}

\begin{figure}[H]
\centering
\includegraphics[width=0.6\columnwidth]{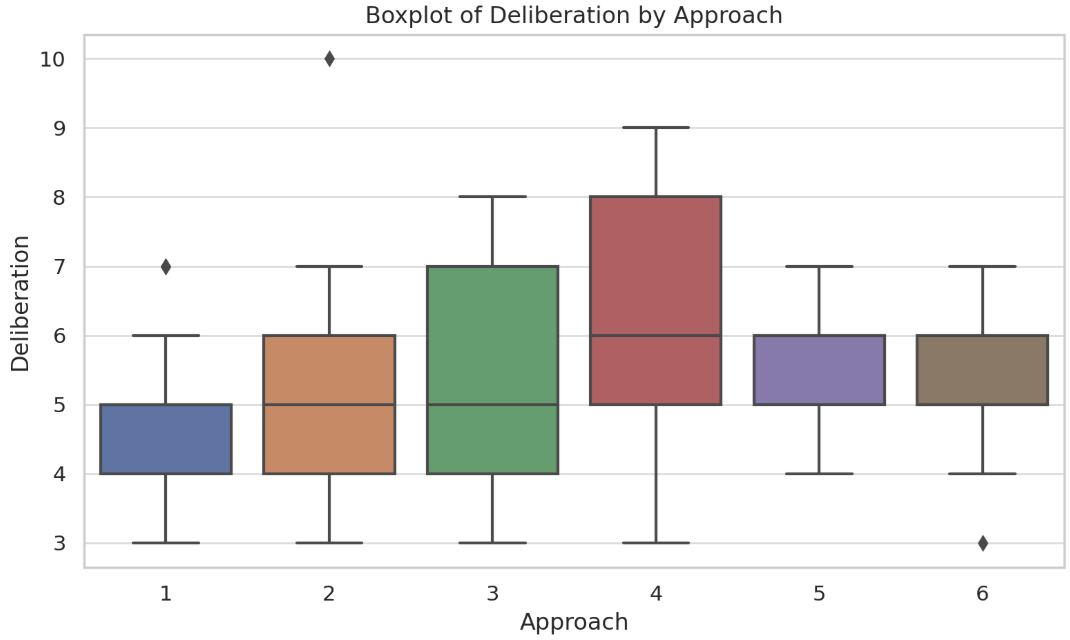}
\caption{BoxPlot Deliberation Times based on six selected languages.}
\label{fig:boxPlotByDeliberation}
\end{figure}

Figures \ref{fig:boxPlotByTotalTime},\ref{fig:boxPlotByPlanWriting}, and \ref{fig:boxPlotByDeliberation}  depict box plots for total development, plan writing and deliberation times, respectively, for six languages. The languages are enumerated as: 1) Jason Fuzzy Int., 2) Jason Rule Fuzzy, 3) Jason Bool, 4) Jade, 5) Java, 6) C++.

\subsection{Group-wise Results}
Table \ref{tab:EmpiricalPerformanceDevelopmentGroups} displays performance statistics of three groups which developed the given three case studies in two different programming languages.

\begin{table}[H]
\centering
\scriptsize
\resizebox{\textwidth}{!}{%
\begin{tabular}{|l|
                |c|c|c|c|c
                |c|c|c|c|c
                |c|c|c|c|c
                |c|c|c|c|c|}
\hline
\textbf{Name} & 
\multicolumn{5}{c|}{\textbf{Total Time}} & 
\multicolumn{5}{c|}{\textbf{Plan Writing}} & 
\multicolumn{5}{c|}{\textbf{Deliberation}} & 
\multicolumn{5}{c|}{\textbf{Debug}} \\
\hline
 & Mean & Std & Min & Max & Cnt. 
 & Mean & Std & Min & Max & Cnt. 
 & Mean & Std & Min & Max & Cnt. 
 & Mean & Std & Min & Max & Cnt. \\
\hline
Fuzzy Agents (G1) & 28.48 & 8.55 & 11 & 52 & 42 & 16.07 & 5.88 & 8 & 35 & 42 & 5.14 & 1.34 & 3 & 10 & 42 & 2.81 & 1.37 & 1 & 6 & 42 \\
\hline
Bool Agents (G2)  & 32.79 & 9.27 & 12 & 47 & 42 & 15.12 & 5.49 & 6 & 27 & 42 & 5.69 & 1.89 & 3 & 9 & 42 & 2.57 & 1.15 & 1 & 5 & 42 \\
\hline
Embeddeds (G3)    & 28.48 & 7.73 & 12 & 39 & 42 & 15.98 & 4.48 & 8 & 25 & 42 & 5.57 & 1.06 & 3 & 7 & 42 & 2.21 & 1.02 & 1 & 5 & 42 \\
\hline
\end{tabular}
}
\caption{Performance statistics (mean, std, min, max, count) across language level groups and development phases.}
\label{tab:EmpiricalPerformanceDevelopmentGroups}
\end{table}

Table \ref{tab:AnovaResultsGroup} displays the carried out ANOVA t-test for group differences having a significance degree at p$<$0.05 for total development, plan writing, deliberation and debugging phases.



\begin{table}[ht]
\centering
\small
\caption{ANOVA for Group Differences (significance at p$<$0.05)}
\begin{tabular}{|l|c|c|}
\hline
\textbf{Metric} & \textbf{p-value} & \textbf{Significant?} \\
\hline
Total Time      & 0.031            &  Yes       \\
Plan Writing    & 0.665            &  No        \\
Deliberation    & 0.204            &  No        \\
Debug          & 0.073            & Borderline           \\
\hline
\end{tabular}
\label{tab:AnovaResultsGroup}
\end{table}

Tukey HSD test was performed after significant ANOVA results to identify which specific groups differ from each other. While ANOVA confirms that a difference exists, it does not indicate which groups the difference lies between. Tukey provides pairwise comparisons and adjusts for multiple testing to avoid false positives. This study revealed that specific approaches (e.g., Language Java vs. Language Jason Fuzzy-BDI Integrated ) significantly differed in total execution time. This post-hoc analysis was essential to interpret the source and direction of group-level differences with statistical confidence.

Table \ref{tab:GroupTotalTimeTukey} indicates the results for Group-level total time that none of the group comparisons reached statistical significance at the p$<$0.05 level. The comparisons between Group 1 and 2, and between Group 2 and 3, both have p-values of 0.0577, which are borderline but not significant. This suggests that, while Group 2 may have taken slightly more time on average, at the group level, total development time does not differ significantly.

\begin{table}[ht]
\centering
\small
\caption{Group - Total Time Multiple Comparison of Means (Tukey HSD, FWER=0.05)}
\begin{tabular}{|c|c|r|c|r|r|c|}
\hline
group1 & group2 & \multicolumn{1}{c|}{meandiff} & p-adj & \multicolumn{1}{c|}{lower} & \multicolumn{1}{c|}{upper} & reject \\
\hline
G1 & G2 & 4.3095 & 0.0577 & -0.111 & 8.7301 & False \\
G1 & G3 & 0.0    & 1.0    & -4.4206 & 4.4206 & False \\
G2 & G3 & -4.3095 & 0.0577 & -8.7301 & 0.111  & False \\
\hline
\end{tabular}
\label{tab:GroupTotalTimeTukey}
\end{table}

\begin{figure}[H]
\centering
\includegraphics[width=0.6\columnwidth]{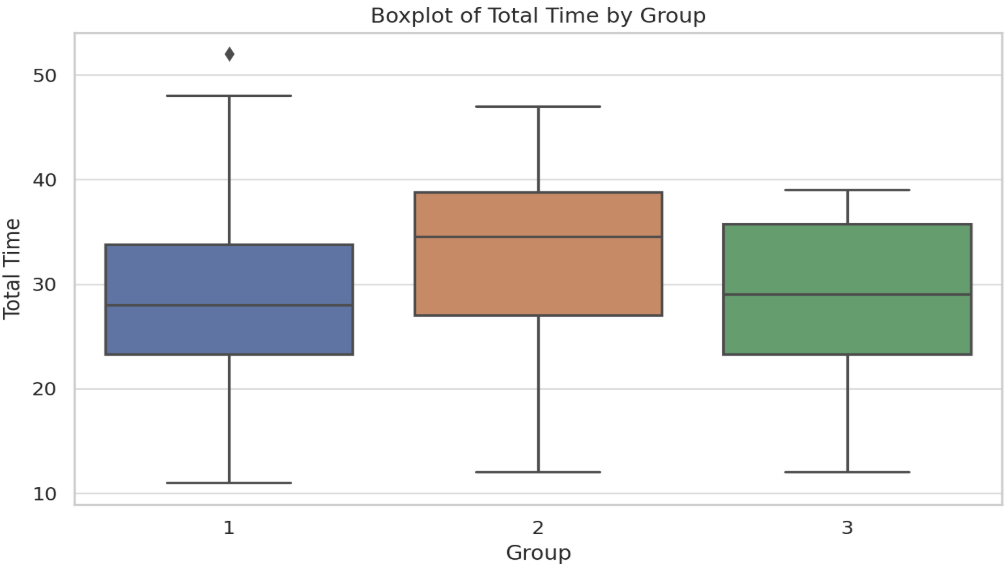}
\caption{BoxPlot Total Times based on three evaluation groups.}
\label{fig:boxPlotByTotalTimeGroup}
\end{figure}
\begin{figure}[H]
\centering
\includegraphics[width=0.6\columnwidth]{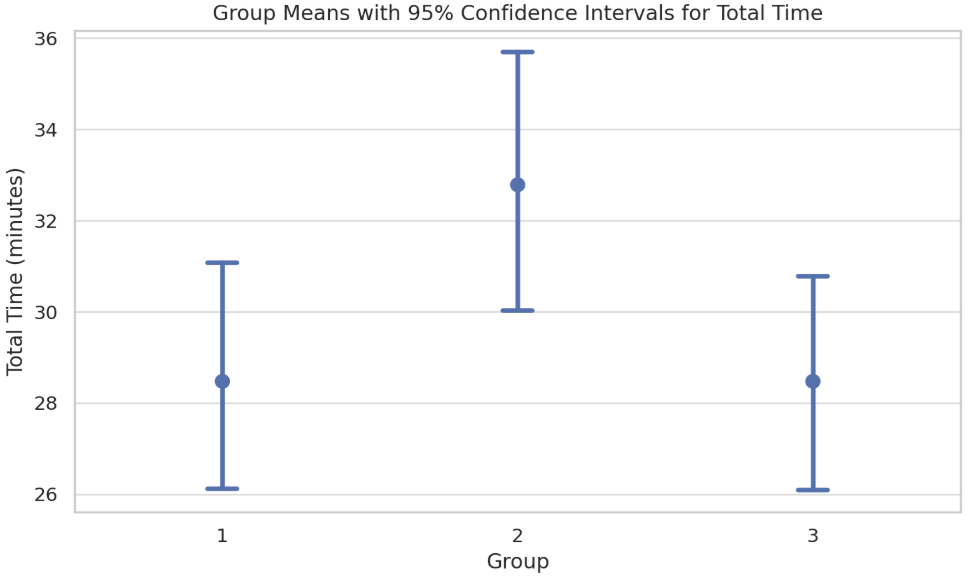}
\caption{Confidence Intervals for Total Development Time based on evaluation groups.}
\label{fig:ConfidenceIntervals95}
\end{figure}

Figure \ref{fig:boxPlotByTotalTimeGroup} shows a box plot for total times on three evaluation groups. Lastly, Figure \ref{fig:ConfidenceIntervals95} shows group-wise confidence intervals with \%95 for total development time.  In the following section, the paper is discussed, including the reported results.


\section{Discussions}

 This paper presents an empirical and experimental evaluation of programming languages starting from traditional OOP to hybrid BDI frameworks. The empirical work here provides a basis to understand how various levels of abstraction in programming influence both development effort and performance of execution for developing intelligent systems. By comparing and contrasting procedural, object-oriented, and agent-oriented paradigms across various application contexts, the results establish that there can be no universal programming technology. Programming languages such as C++ ensure deterministic performance, a necessity in hard real-time systems, lacking further cognitiveness.

On the other hand, agent-oriented programming languages such as Jason and fuzzy logic extensions enable high-level specification of behaviour to perform proactivity and uncertainty handling at the expense of increased run-time overhead. Specifically, when BDI agents and overall empirical evaluation are considered, referring to the embedded operating systems \citet{hahm2020reliable}, it is expected that hard real-time response must be between 50 microseconds and 10 milliseconds for reliability.  In addition, a study by \citet{donga2022critical} determines the operational range of critical soft real-time systems to be 10 ms to 100 ms. This may make Jason-BDI-based systems more suitable for deadline-tolerant applications when the recursive plan branches, i.e., recursive calls are not overused. Subjectively, it should not exceed 5 calls to make it under 100ms (see Figure \ref{fig:RecursiveLinePlot}) or at most 10 calls to keep it around 100ms. Jason-based approaches require these recursive plan calls to perform proactive behaviour as degree of intelligence. Nevertheless, it should also be noted that achieving a real-time agent system remains an open gap \citet{calvaresi2021real}. Eventually, these agent-oriented approaches enable autonomy, social ability, and decision-making in uncertain environments, all of which are essential in current cyber-physical and intelligent control systems and distributed information processing.

A second, closely related and equally important concept is Physical AI. Physical AI deals with the embodiment of intelligence in physical agents such as mobile robots, autonomous drones, or human-interactive machines. In these contexts, intelligent reasoning must be tightly integrated with physical sensing and actuation. The architecture must support both high-level planning and low-level control, and must include safety, responsiveness, and adaptability. The conclusions drawn from this study are that agent-oriented reasoning models, particularly those based on fuzzy logic, provide a natural fit for layered control architectures. Such architectures must be able to respond rapidly to environmental stimuli, as well as provide the flexibility to implement deliberative actions on time scales that are longer than the response. Future innovations may involve hybrid architectures that combine symbolic agents with language model guidance and real-time control layers, enabling direct physical interaction with the world while considering short-, mid-, and long-term system objectives.

Overall, the results of our empirical and WCET-based evaluations reveal clear trade-offs between abstraction level and system efficiency in IoT and CPS. As anticipated, languages with higher abstraction, such as Jason, Jade, and especially fuzzy BDI extensions, incur increased worst-case execution times, particularly under recursive and branching logic scenarios. However, these platforms significantly reduce development time and cognitive load, especially in complex, rule-driven control applications like those in our Cleaning Robots and Smart Production Line case studies.
This finding aligns with previous studies (e.g., \citet{gavigan2024quantifying} ; \citet{becker2025expedited}), which also observed that agent-based models offer enhanced maintainability and reasoning capacity at the cost of timing precision. Our study extends these insights by including fuzzy-BDI frameworks, demonstrating that fuzzy reasoning mechanisms retain the expressiveness of symbolic agents while introducing only moderate additional computational overhead.
We also observed that object-oriented languages such as Java and Jade serve as a middle ground: they strike a balance between developer productivity and runtime performance. These results support \citet{jordan2015feature}, who argued that object-oriented paradigms offer modularity without the high reasoning overhead of fully deliberative systems.
Finally, the statistical analyses—using ANOVA and post hoc Tukey HSD tests confirm that the observed trends in WCET and development time are significant across hardware types (PC and Raspberry Pi 3) and participant profiles. This empirical robustness positions our study as a practical reference for selecting programming languages in real-time, agent-driven CPS.

\subsection{Execution Time}

The empirical findings reveal definite and consistent performance differences between agent implementations along the dimensions of underlying programming language and architectural style. Jason-based agents, and especially those using fuzzy logic, registered the longest overall execution times across all measured dimensions when recursiveness increased to 36 calls. Such high values, displaying averages of over 100 milliseconds and maximums of over 500 milliseconds, highlight the computational overhead of incorporating sophisticated reasoning mechanisms into agent design. In comparison, conventional imperative-style implementations, such as C++, Java, and Jade, recorded considerably shorter total execution times, typically below 30 milliseconds. This indicates that simpler control flows and the absence of deliberative overhead enable more efficient execution in time-critical settings. However, having 36 recursive calls and hitting 555ms is relatively poor agent-programming convention or a disrespect for complex systems based on limited hardware capacity. With or without fuzzy enhancement, BDI reasoning creates a burden for run-time performance if recursive plan calls are coded spontaneously. Fortunately, there are architectural enhancements for safety-critical systems \citet{becker2025expedited,boissier2020multi}.

In measuring planning behaviours, loosely coupled fuzzy-BDI agents had the highest mean plan writing time, indicating that writing extra belief-rules creates additional development complexity, as expected. Jade agents also had high plan-writing times, which might be because it is an extension of Jade and requires extra syntactic burden as measured in \citet{karaduman2023rational}. Surprisingly, however, integrated fuzzy-BDI agents, although having the highest overall total execution time, did not dominate plan-writing times. This indicates that the majority of their computational overhead is factored elsewhere in the reasoning process, perhaps in internal rule evaluation or fuzzy inference phases. However, this may be reduced by applying optimisations.

\subsection{Development Efficiency}

The regression analysis conducted on the collected data demonstrates several statistically significant effects influencing total task completion time during programming language-based coding activities. Planning time emerged as a strong predictor ($\beta$ = 0.65, p $<$ 0.001), indicating that increased allocation to planning and algorithmic reasoning significantly extends overall task duration. This finding aligns with established cognitive theories suggesting that pre-coding strategising imposes a measurable temporal cost, which should be balanced against execution efficiency.

Group membership was also a significant determinant: participants in Group 2 required on average 3.16 additional minutes to complete their tasks compared to the baseline Group 1 (p $<$ 0.001). This disparity may be attributed to differences in expertise, familiarity with the programming approach, or intra-group problem-solving dynamics, underscoring the importance of considering participant characteristics in performance assessments.

Similarly, the choice of programming approach influenced task duration notably. Jade accounted for a 2.93-minute increase relative to other approaches (p $<$ 0.001), suggesting that this method encompasses greater syntactic complexity or abstraction, thereby demanding more processing time. This result has direct implications for selecting programming languages or frameworks in both educational and industrial contexts.

Regarding the experimental setting, Cases 2 and 3, Cleanig Robots and Smart Production Line, respectively, resulted in significantly longer completion times than Case 1, with increases of 7.02 and 8.73 minutes (p $<$ 0.001). These substantial differences likely reflect increased problem complexity or debugging difficulty inherent to these scenarios, highlighting the critical role of task design in evaluating programming performance metrics.

Importantly, interaction terms between variables largely failed to reach significance (p $>$ 0.05), indicating that the main effects of planning time, group, approach, and case largely operate independently without substantial synergistic interactions. This simplifies the interpretability of the model and supports the robustness of additive predictors in this context.

The model’s overall explanatory power was high (R² = 0.752), accounting for more than three-quarters of the variance in total task time. This strong fit validates the empirical approach and reinforces confidence in the identified factors as primary contributors to temporal variation in programming tasks.

\subsection{Threats to the Validity}

This subsection discusses threats to validity, focusing on potential limitations or elements that could compromise the integrity of the experiments and the results obtained. These threats highlight areas where bias or inaccuracy could potentially impact the study’s outcomes. By identifying these concerns, researchers aim to provide a transparent view of the study’s limitations and their potential impact on the validity of the findings.

\subsubsection{Conclusion Validity}

Within the conclusion validity, the interest lies in checking the relationship among the subjects (experimental units, participants) and the obtained results. The goal is to establish a statistical relationship that corroborates the observations and hypotheses at a statistically significant level. To accomplish this, a series of scenarios was obtained from various experiments and participants performed on the systems.

The statistical tests applied in these experiments were chosen with care to provide accurate and trustworthy results. Moreover, complex, modular, and heterogeneous case studies from the literature were selected and extended to provide a flexible and comprehensive platform for conducting various experiments at different stages. 

The number of subjects (n = 21) was small but sufficient to detect repeated trends at various abstraction levels. When similar studies \citep{challenger2016systematic,kardacs2017supporting,kardas2018domain} are examined, it becomes clear that MAS studies have generally involved limited participant numbers. Therefore, this situation narrows down the size of potential participants. Groups were not assigned randomly but constituted based on a priori expertise levels (i.e., AOP and OOP experience), which can lead to selection bias. However, cross-comparison of several languages within and between groups mitigated the effect of individual variance. All code was assessed on the same criteria and anonymised to prevent evaluator bias.

This comprehensive setup enables the investigation of various system attributes and variables, allowing researchers to conduct detailed analyses and draw significant conclusions. Through thoughtful experiment design, the selection of appropriate statistical tests, and intensive case studies, the research aims to establish a robust foundation for determining the relationship between the subjects and the results obtained. This adds to the conclusion validity of the research and enhances the reliability of the findings.

\subsubsection{Internal Validity}

Internal validity refers to the extent to which a causal relationship between the treatment and the observed result can be confirmed, ensuring that outcomes are not influenced by uncontrolled or external factors. While participants were randomly assigned to balanced groups and provided with equal instructional materials, threats to internal validity may still arise due to differences in individual coding habits, interpretation of requirements, or engagement with the task. To address these concerns, all participants were given the same code templates, watched identical instructional videos, and were allocated the same time limits. A 10-day break between sessions was also included to minimise learning transfer and fatigue, although it may not completely eliminate residual memory effects.

In addition to these precautions, two independent variables were held constant throughout the study. These were the reactive and proactive scenario features, which consisted of predefined goals that all participants were required to achieve. At the start of the experiment, participants were not informed about the specific features they would be implementing, reducing any preparation-related bias.

Each participant completed the experiment only once per language and had no prior exposure to the same case studies. Participants worked in parallel using different programming languages, which enabled cross-validation. Communication between participants was strictly prohibited, and both remote and in-person sessions were closely monitored to maintain experimental integrity.

Because of the limited number of participants, a within-subjects design was used. Each participant completed tasks in two different programming environments. A consistent time interval was maintained between sessions to reduce carry-over effects. Finally, case studies were carefully selected based on previous literature to reflect realistic industrial conditions. The selection of the case studies was not random. These case studies were curated from the studies in the literature \citet{karaduman2023rational,karaduman2024impact,weyns2020introduction,becker2025expedited}.

\subsubsection{Construct Validity}
Construct validity concerns the generalisation of results based on the conducted experiments.
In this research, all of the languages and WCET measurement approaches and comparison methods are rational approaches that have been the subject of intense study for decades. To minimise the risk of obtaining results by chance and to ensure the validity of our findings, we conducted multiple repetitions of our experiments, as mentioned in \citet{miller2021analysis,miller2021performance}. The measurements were taken according to high-quality sampling in the system, which we also sought to strengthen. Thus, generalisation is not likely to detract from construct validity.

No information about the research has been provided to the participants, and no discussion about the pros and cons of the selected languages has been conducted. Therefore, the participants cannot predict the expected results of this study. As a result, it is seen as a
minor threat to construct validity. Both agent-oriented programming languages and OOP ones conform to the same scenario and their requirements. Therefore, the generalisation
of it cannot have an adverse effect.

Overall, we concentrated on two major measures: WCET  and Development Time, as surrogates for engineering effort and performance. These do not account for additional software qualities, such as maintainability, readability, or scalability. Although we chose these measures because they are of utmost significance in embedded, IoT and CPS environments for controlling the system, future research may include a broader range of software quality attributes.

\subsubsection{External Validity}

External validity refers to the extent to which the findings of a study can be generalised to real-world situations or other contexts beyond specific experimental conditions. In our study, we took measures to address potential threats to external validity and ensure the relevance and applicability of our experiment results to software engineering techniques.  One common threat to external validity is the reliance on PhD and master's students as participants. In this study, that risk was reduced by carefully selecting participants who shared comparable backgrounds and had relevant industry/programming experience.

The generalizability of the findings could be limited by the specific technologies, case studies, and participant demographics used. Although the chosen programming environments (C++, Java, Jade, Jason, Jason fuzzy-BDI) cover a spectrum of abstraction levels and reasoning paradigms, they might not be representative of all software engineering settings, particularly in application domains outside CPS, IoT, robotics and embedded systems. Additionally, empirical tasks were limited to rule/plan/condition-based control problems with extra minor requirements, which could restrict applicability to more general software development contexts.






\section{Conclusion and Future Work}


This paper presents a comprehensive, empirical comparison of six programming platforms—spanning imperative (C++), object-oriented (Java), agent-oriented (Jade, Jason), and fuzzy-BDI paradigms (two fuzzy-BDI Jason extensions) across three case studies targeting real-time IoT and CPS applications. By jointly evaluating development effort and WCET, this study provides a dual-lens perspective on how abstraction level impacts both software quality and runtime feasibility. Our findings suggest that while low-level languages such as C++ offer superior timing performance, high-level agent-based approaches like Jason and its fuzzy extensions significantly reduce development complexity and promote behavioural modularity. The results confirm the well-established abstraction-efficiency trade-off but also highlight that fuzzy reasoning frameworks can achieve expressive intelligence with modest performance overheads. These findings are especially relevant for researchers and developers in embedded systems and cyber-physical domains, particularly in safety-critical environments where balancing reasoning capacity and actuation latency is essential. In systems where context-awareness, autonomous decision-making, or inter-agent cooperation is required, agent-oriented paradigms provide benefits in terms of modularity, transparency, and scalability. Furthermore, incorporating fuzzy reasoning techniques enhances system robustness in uncertain or dynamically changing environments. Moreover, the use of multiple case studies and statistical rigour adds robustness to the findings, making them relevant not only for researchers and developers but also for system architects and educators, too.

The current fuzzy-BDI framework relies on type-1 fuzzy sets, which limits the expressiveness of the reasoning process under higher-order uncertainty. There is no automated mechanism for generating fuzzy rules, and the manual design of rule sets remains time-consuming and prone to errors. Future work will explore the integration of type-2 fuzzy sets, which provide an additional level of uncertainty modelling by introducing uncertainty within the membership functions themselves. This added flexibility enables better handling of ambiguous or imprecise information, thereby improving the quality of decision-making processes.

In addition to improved reasoning mechanisms, learning-based techniques can support the automated generation and adaptation of fuzzy membership functions, reducing the development burden and increasing adaptability over time \citet{bosello2019programming,parmiggiani2025together}. Finally, the introduction of domain-specific modelling languages can further lower development barriers. These modelling tools make it easier for domain experts to participate in the system design process without requiring deep programming expertise \citet{karaduman2024dsml4jacamo}, applying model-driven engineering. These developments align with the goals of industrial information integration by supporting the design of intelligent, flexible, and explainable systems that can operate in complex industrial environments.







\bibliographystyle{apalike}
\bibliography{bibtex} 

\end{document}